%% file: VLFBBC_v20.tex
\documentclass[onecolumn]{IEEEtran}

\input{preamble}
\usepackage{xspace}
\usepackage[ colorlinks = true,
             linkcolor = blue,
             urlcolor  = blue,
             citecolor = red,
             anchorcolor = green,
]{hyperref}

\newcommand{\upC}{\overline{C}}
\usepackage{cite}
\allowdisplaybreaks[2]
\begin{document} 

%\title{On Gaussian Channels with Variable-Length Feedback and Non-Vanishing Error Probabilities} 
%\title{On AWGN Channels and Gaussian MACs with Variable-Length Feedback and Non-Vanishing Error~Probabilities} 
\title{On the Reliability Function of the Common-Message Broadcast Channel with Variable-Length Feedback} 
%\author{\IEEEauthorblockN{Lan V. Truong, and Vincent Y.~F.~Tan}\\
%\IEEEauthorblockA{Department of Electrical and Computer Engineering\\
%National University of Singapore \\ Emails: \url{lantruong@u.nus.edu}; \url{vtan@nus.edu.sg}}}

\author{Lan V.\ Truong  $\,$  and $\,$
        Vincent Y.~F.~Tan
\thanks{The authors are with the  Department of Electrical and Computer Engineering, National University of Singapore (NUS). V.~Y.~F.~Tan is also with the Department of Mathematics, NUS. Emails: \url{lantruong@u.nus.edu};  \url{vtan@nus.edu.sg}} \thanks{The authors are supported by an NUS Young Investigator Award (R-263-000-B37-133) and a  Singapore Ministry of Education (MOE) Tier 2 grant (R-263-000-B61-112).}
}
%%%%%%%%%%%%%%%%%%%%%%%%%%%%%%%%%%%%%%%%%%%%%%%%%%%%%%%%%%%%%%
%%%%%%%%%%%%%%%%%%%%%%%%%%%%%%%%%%%%%%%%%%%%%%%%%%%%%%%%%%%%%%

\maketitle
 
\begin{abstract}  We derive upper and lower bounds on the reliability function for the common-message discrete memoryless broadcast channel   with  variable-length feedback. We show that the bounds are tight when the broadcast channel is stochastically degraded. For the achievability part, we adapt Yamamoto and Itoh's coding scheme by controlling the expectation of the maximum of a set of stopping times.  For the converse part,  we adapt Burnashev's proof techniques for establishing the  reliability functions for (point-to-point) discrete memoryless channels with variable-length feedback and   sequential hypothesis testing. 
\end{abstract}   
 
\section{Introduction}
Shannon~\cite{Sha56} showed that noiseless feedback does not increase the capacity of single-user memoryless channels. Despite this seemingly negative result, feedback  significantly simplifies coding schemes and improves the performance in terms of the error probability~\cite{SK66, YHK2011a, Wu14, Burn14,TFT16}. Burnashev~\cite{Burnashev1976} demonstrated that the {\em reliability function} for the discrete memoryless channel (DMC) with feedback improves dramatically when the transmission time is random. This is known as {\em variable-length feedback}. In fact, the reliability function of a DMC with variable-length feedback   admits a particularly simple expression
\begin{align}
E(R)= {B_1} \left(1-\frac{R}{C}\right)\label{eqn:burn}
\end{align} for all rates $0\le R\le C$, where $C$ is the capacity of the DMC and $B_1$ is determined by the relative entropy between conditional output distributions of the two most ``most distinguisable'' channel input symbols~\cite{Burnashev1976}. Yamamoto and Itoh~\cite{YamamotoItoh1979} proposed a simple and conceptually important two-phase coding scheme that attains the reliability function in \eqref{eqn:burn}. Since these  reliability function (or error exponent) results are of paramount importance in practical single-user feedback communication systems, we are motivated to extend the results to a simple network scenario---namely, the discrete memoryless broadcast channel (DM-BC) with a common message (also known as the common-message DM-BC) \cite{Wu14,Trillingsgaard15,Trillingsgaard2016}. We provide upper and lower bounds on the reliability function and show that the bounds coincide  if the DM-BC is stochastically degraded. In this scenario, the reliability function is dominated by the ``worst branch'' of the DM-BC. 
\subsection{Main Contributions}\label{sec:main_contr}
Our main technical contributions are as follows:
\begin{itemize}
\item Firstly, for the achievability part, we generalize Yamamoto and Itoh's coding scheme~\cite{YamamotoItoh1979} so that it is  applicable to  the DM-BC with a common message  and variable-length feedback.  In this enhanced   scheme, we supplement some new elements to the original   arguments in~\cite{YamamotoItoh1979}. These include (i) defining an appropriate  set of $K$ stopping times and (ii) proving that the expectation of the maximum of these $K$ stopping times  can be appropriately bounded assuming that the individual stopping times' expectations and variances are also appropriately bounded. This complication of having to control the {\em maximum} of a set of stopping times does not arise in single-user scenarios such as~\cite{Burnashev1976,Burnashev80,Nak08}.  
\item Secondly, for the converse part, we adapt and combine       proof techniques  introduced by Burnashev for two different problems---namely, the reliability function for DMCs with variable-length feedback in~\cite{Burnashev1976} and  that for sequential hypothesis testing in~\cite{Burnashev80}. This allows us  to obtain an upper bound for the reliability function for the common-message DM-BC with variable-length feedback. There is an alternative  and more elegant  proof technique to establish the converse part of~\eqref{eqn:burn} by Berlin {\em et al.}~\cite{Berlin2009a} but generalizing the technique therein to our setting does not seem to be feasible. %Instead we combine Burnashev's original ideas for DMCs with variable-length feedback in~\cite{Burnashev1976} and sequential hypothesis testing in~\cite{Burnashev80} to establish a meaningful upper bound on the reliability function. 
%  that does not lead to 
%since a direct application of proof techniques~\cite{Burnashev80},~\cite{Berlin2009a} does not lead to the same result. By adapting the proof techniques in~\cite{Burnashev1976} and~\cite{Burnashev80}, we also make the converse proof simpler. 
\item Thirdly, even though the bounds on the reliability function do not match for general DM-BCs, we   identify a particular class of DM-BCs, namely {\em stochastically degraded} DM-BCs~\cite[Sec.~5.6]{elgamal} for which the reliability function  is known exactly. For the {\em less capable} DM-BCs (to be defined formally in Definition~\ref{def:order}), even though we only have bounds on the reliability function, from these bounds, we can establish the   capacity of such channels with variable-length feedback. %Identify a class of orderings of channels such as less capable, stochastically degraded for which the reliability function is known exactly.
%\item Obtain the exact reliability function for compound channel with variable-length feedback.
\end{itemize}
\subsection{Related Works}
We summarize some related works in this subsection. In~\cite{Burnashev80}, Burnashev extended the ideas in his original paper in DMCs with variable-length feedback~\cite{Burnashev1976} to be amenable to the more general problem of sequential hypothesis testing. In particular, he studied the minimum expected number of observations (transmissions) to attain some level of reliability and found the  reliability function for large  class of single-user channels (beyond DMCs), including the Gaussian channel~\cite{Burnashev80}. Berlin {\em et al.}~\cite{Berlin2009a} provided a simple converse proof for Burnashev's reliability function~\cite{Burnashev1976}. Their converse proof suggests that a {\em communication} and a {\em confirmation} phase are implicit in any scheme for which the probability of error decreases exponentially fast with (the optimal) exponent given by \eqref{eqn:burn}. Under this viewpoint, this converse proof approach is parallel to the Yamamoto and Itoh's achievability scheme~\cite{YamamotoItoh1979}. Nakibo\u{g}lu and Gallager~\cite{Nak08} investigated variable-length coding schemes for (not necessarily discrete) memoryless channels with   variable-length feedback  and with cost constraints and established the reliability function. %They showed that this function is concave in the pair $(R, P)$, where $P$ is average energy constraint, and generalized the linear reliability function of Burnashev~\cite{Burnashev1976} to include cost constraints. 
Their achievability proof   is an extension of Yamamoto and Itoh's~\cite{YamamotoItoh1979} and their  converse proof uses two bounds on the difference of the conditional entropy random variable similarly to~\cite{Burnashev1976} with some extra arguments to account for the average cost constraints. Chen, Williamson, and Wesel~\cite{TYChen2014} proposed a two-phase stop-feedback coding scheme where each phase uses an incremental redundancy scheme achieving Burnashev's reliability function~\eqref{eqn:burn} while maintaining an expansion of the size of the message set that yields a small backoff from capacity. Their coding scheme uses a stop-feedback code~\cite{Yury2011} for the first-phase  and a  sequential probability ratio test~\cite{Wald1947} for the second-phase.

We  also mention the work by Shrader and Permuter~\cite{Brooke2009a} who studied the feedback capacity of compound channels~\cite{Wolfowitz59, Blackwell59}. The authors considered  fixed-length feedback while our focus is on variable-length feedback. Mahajan and Tatikonda~\cite{Mahajan12} considered the variable-length case for the same channel and established inner and outer bounds on the so-called error exponent region. While the common-message DM-BC we study is somewhat similar to the compound channel~\cite{Wolfowitz59, Blackwell59}, the techniques we use are different and we  establish  the {\em exact} reliability function for stochastically degraded DM-BCs.  Tchamkerten  and Telatar,  in a series of elegant works~\cite{Tchamkerten05, Tchamkerten06, Tchamkerten06a}, considered conditions in which one can achieve Burnashev's exponent in \eqref{eqn:burn} universally, i.e., without precise knowledge of the DMC.

Recently, there have also been numerous efforts to establish fundamental limits of single- and multi-user channels with variable-length feedback for {\em non-vanishing} error probabilities. See~\cite{Yury2011,Trillingsgaard:2014,Trillingsgaard15,Trillingsgaard2016,TruongTan16a} for an incomplete list. However, we are concerned with quantifying the exponential rate of decay of the error probability similarly to~\eqref{eqn:burn}. 

\subsection{Paper Organization}
The rest of this paper is structured as follows: In Section \ref{sec:dmc_bc}, we provide the problem formulation  for the DM-BC with a common message under variable-length feedback with termination. The main results concerning the reliability function,  conditions under which the results are tight, and some accompanying discussions are stated in Section~\ref{sec:mainresult}. In Section~ \ref{achproof}, we provide the achievability proof. The converse proof is provided in Section~\ref{sec:conveseproof}. We also explain the novelties of our arguments relative to existing works at the end of the proofs in Sections~\ref{achproof} and~\ref{sec:conveseproof}. Auxiliary technical results that are not essential to the main arguments are relegated to the appendices. 
\section{Problem Setting}\label{sec:dmc_bc}
\subsection{Notational Conventions} We use asymptotic notation such as $O(\cdot)$ in the standard manner; $f(n)=O(g(n))$ holds if and only if the implied constant $\limsup_{n\to\infty} |f(n)/g(n)|<\infty$. Also $f(n)=o(g(n))$ if and only if $\lim_{n\to\infty}|f(n)/g(n)|=0$. In this paper, we use $\ln x$ to denote the natural logarithm so information units throughout are in nats. The binary entropy function is defined as $h(x):=-x\ln x-(1-x)\ln(1-x)$ for $x\in [0,1]$. We also define the function $(x)_a:=x \bone\{x\geq a\}$ for  $x, a\in \bbR$.  The minimum of two numbers $a$ and $b$ is denoted interchangeably as $\min\{a,b\}$ and $a\wedge b$. As is usual in information theory $Z_i^j$ denotes the vector $(Z_i,Z_{i+1},\ldots, Z_j)$. %For a sequence $\{A_n\}_{n=1}^{\infty}$, we denote by notation $Y_{j,p}^q:=(Y_{j,p}, Y_{j,p+1}, \cdots, Y_{j,q})$ and $Y_j^q=Y_{j,1}^q$.

For any discrete product sample space $\calZ \times \calT$, a sigma-algebra $\calF$ on $\calZ \times \calT$, two random variables $Z, T$ (not necessary measurable with respect to $\calF$), and two regular conditional probability measures $\bbP(\cdot|\calF), \bbQ(\cdot|\calF)$ on $\calZ \times \calT$, define  
\begin{align}
\label{eq1notation}
 \calH(Z|\calF)&:=-\sum_{z\in \calZ} \bbP (z|\calF)\ln \bbP(z|\calF),\\
 H(Z) &:=\calH(Z|\sigma(\emptyset,\calZ\times \calT)),\\
D(\bbP\|\bbQ)&:=\sum_{(z,t) \in \calZ\times \cal T} \bbP(z,t|\sigma(\emptyset,\calZ\times \calT)) \ln \frac{\bbP(z,t|\sigma(\emptyset,\calZ\times \calT))}{\bbQ(z,t|\sigma(\emptyset,\calZ\times \calT))},\\ 
 %\calD(\bbP\|\bbQ|\calF) &: =\sum_{(z,t) \in \calZ\times \cal T} \bbP(z,t) \ln \frac{\bbP(z,t)}{\bbQ(z,t)},\\ 
 \calI(Z;T|\calF)&:=\sum_{(z,t) \in \calZ\times \cal T} \bbP(z,t|\calF) \ln \frac{\bbP(z,t|\calF)}{\bbP(z|\calF)\bbP(t|\calF)},\\ 
%&=\sum_{(z,t)\in \calZ \times \calT} \bbP(Z=z,T=t|\calF)\ln \frac{\bbP(Z=z,T=t|\calF)}{\bbP(Z=z|\calF) \bbP(T=t|\calF)},\\
 I(Z;T)&:=\calI(Z;T|\sigma(\emptyset,\calZ \times \calT). \label{eq6notation}
\end{align}
If $\calF=\sigma(Y^n)$ for some vector $Y^n$, we write $\sigma(Y^n)$ as $Y^n$ in all above notations~\eqref{eq1notation}--\eqref{eq6notation}  for simplicity~\cite{Billingsley}.

\subsection{Basic Definitions}
\begin{definition}
\label{def1}
A $(M,N)$-{\em variable-length feedback code with termination (VLFT)} for a $K$-user DM-BC   $P_{Y_1,Y_2,\ldots,Y_K|X}$  with a common message, where $N$ is a positive real and $M$ is a positive integer, is defined by
\begin{itemize}
\item A set of equiprobable messages $\calW=\{1,2,\ldots,M\}$.
\item A sequence of encoders $f_n: \calW \times \calY_1^{n-1}\times \calY_2^{n-1} \times\cdots \times \calY_K^{n-1} \to \calX, n\geq 1$, defining channel inputs
\begin{align}
X_n=f_n(W,Y_1^{n-1},Y_2^{n-1},\cdots,Y_K^{n-1}).\label{eqn:enc}
\end{align} 
\item $K$ sequences of decoders $g^{(j)}_n: \calY_j^n \to \calW, j=1,2,\ldots,K$, providing the best estimate $W$ at time $n$ at the corresponding decoders.
\item A stopping random variable $\tau:=\max\{\tau_1,\tau_2,\ldots,\tau_K\}$, where for each  $j\in \{1,2,\ldots,K\}$, $\tau_j$ is a stopping time of the filtration $\{\sigma(Y_j^n)\}_{n=0}^{\infty} $. Furthermore, $\tau$ satisfies the following constraint:
\begin{align}
\label{eq8def}
\bbE(\tau) \leq N.
\end{align}
\end{itemize}
\end{definition}
The final decision at decoder $j = 1,2, \ldots, K$  is computed at time $\tau_j$ as follows:
\begin{align}
\hat{W}_j&=g^{(j)}_{\tau_j}(Y_j^{\tau_j}).
\end{align} 
The {\em error probability} of a given variable-length coding scheme is defined as
\begin{align}
\rvP_{\rme}(R,N):=\bbP\bigg(\bigcup_{j=1}^K \{\hat{W}_j \neq W\}\bigg).
\end{align}
The {\em rate} of the $(M,N)$-VLFT code  (cf.~Definition~\ref{def1}) is defined as
\begin{align}
R_N:=\frac{\ln M}{N}.
\end{align} 
\begin{definition} \label{def:rel}
$(R,E)\in\bbR_+^2$ is an {\em achievable rate-exponent pair} if there exists a family of $(M_N,N)$-VLFT codes (for   $N\to\infty$)  satisfying
\begin{align}
\liminf_{N \to \infty} R_N &\ge R,\label{eq:Rate}\\
\lim_{N\to \infty} \rvP_{\rme}(R_N,N)&=0,\\
\liminf_{N\to \infty}\,\,-\frac{\ln \rvP_{\rme}(R_N,N)}{N} &\geq E,
\end{align}
where $R_N = N^{-1}  \ln M_N$. 
The {\em reliability function} of the DM-BC with VLFT is 
\begin{align}
 E(R):=\sup\{E : (E,R)\mbox{ is an ach.\ rate-exp.\ pair} \}.
\end{align}
\end{definition}
 
In a VLFT code for the DM-BC,  the word ``termination'' is used to indicate that in order to realize the code in a practical setting, one needs to  send a reliable end-of-packet signal by a method  other than using the transmission channel. In other words, the encoder decides when to stop the transmission of signals~\cite{Yury2011,Trillingsgaard2016}.

We now recapitulate a set of orderings of channels~\cite[Ch.~5]{elgamal}.
\begin{definition} \label{def:order}
A DM-BC $P_{Y_1,Y_2,\ldots,Y_K|X}$ is  {\em less capable}\footnote{In the literature~\cite[Sec.~5.6]{elgamal}, the term {\em more capable} is typically used when $Y_1$ is the ``strongest receiver''. However, in our context, $Y_1$ is the ``weakest receiver'' so we use the (somewhat atypical) term {\em less capable}  here.} \cite[Sec.~5.6]{elgamal} (with respect to the first channel $P_{Y_1|X}$) %\red{there is no such thing as less capable! it's less capable ; need  to change ISIT pape} 
if 
\begin{equation}
I(X;Y_1)\le \min_{1\le j\le K} I(X;Y_j)
\end{equation}
 for all $P_X$. A DM-BC $P_{Y_1,Y_2,\ldots,Y_K|X}$ is {\em stochastically  degraded} \cite[Sec.~5.4]{elgamal} (with respect to $P_{Y_1|X}$) if there exists a random variable  $\tilde{Y}_1$ such that 
\begin{align}
 \tilde{Y}_1|\{X=x\} \sim P_{Y_1|X}(\cdot|x), \quad &\forall \,   \tilde{y}_1 \in \calY_1,\quad \mbox{and}\\
 X-Y_j-\tilde{Y}_1, \quad\,&\forall \,  j=2,3,\ldots,K
\end{align}
A DM-BC $P_{Y_1,Y_2,\ldots,Y_K|X}$ is  {\em physically degraded}  \cite[Sec.~5.4]{elgamal}  (with respect to $P_{Y_1|X}$) if 
\begin{equation}
X-Y_j-Y_1
\end{equation}
 forms a Markov chain for all $ j= 2,\ldots,K$.
\end{definition}
Clearly, the set of all physically degraded DM-BCs contained in the set of all stochastically degraded DM-BCs  which is contained in the set of all less capable DM-BCs.   We omit another commonly-encountered set of orderings for DM-BCs, namely {\em less noisy} DM-BCs~\cite[Sec.~5.6]{elgamal}.

\begin{definition} \label{def:info_q} For a DM-BC with a common message and VLFT as in  Definition~\ref{def1}  we define for each $1\leq j \leq K$,
\begin{align}
B&:=\max_{x,x'\in \calX} \min_{1\leq j\leq K} D(P_{Y_j|X}(\cdot|x)\|P_{Y_j|X}(\cdot|x'),\\
B_j&:=\max_{x,x' \in \calX} D(P_{Y_j|X}(\cdot|x)\|P_{Y_j|X}(\cdot|x')),\\
%B_{\mathrm{min}}&:=\min_{1\leq j\leq K} B_j,\\
B_{\mathrm{max}}&:=\max_{1\leq j \leq K} B_j,\\
T_j&:=\max_{x,x'\in \calX, y \in \calY_j}\frac{P_{Y_j|X}(y|x)}{P_{Y_j|X}(y|x')},\\
C&:=\max_{P_X} \min_{1\leq j\leq K} I(X;Y_j),\\
C_j&:=\max_{P_X}I(X;Y_j),\\  
\upC&:=\min_{1\leq j\leq K} \max_{P_X} I(X;Y_j).
\end{align}
\end{definition}

%\red{I find the nomenclature $\upC$ quite counter-intuitive since $\upC\ge C$.  The reason why I used $C$ is because it is capacity of the broadcast channel with fixed-length feedback, so I guess that it is also the capacity of the broadcast channel with variable-length feedback. We also use $B_{\mathrm{min}}$ to express the same. Can you suggest another notation? I have not thought of any better solution yet. You can suggest me to change or help me to change it. Thank you very much.}
\section{Main Results}\label{sec:mainresult}
We now state bounds on the reliability function of the $K$-user DM-BC channel $P_{Y_1,Y_2,\ldots,Y_K|X}$  with a common message and with VLFT.
\begin{theorem}  \label{thm:main_res} For any $K$-user DM-BC channel $P_{Y_1,Y_2,\ldots,Y_K|X}$ with VLFT  (cf.~Definition \ref{def1})  such that $B_{\mathrm{max}}<\infty$, %the following bounds hold
\label{mainthm}
\begin{align}
E(R)&\geq B \left(1-\frac{R}{C}\right),\quad  \forall R<C, \label{eq1:mainthm}
\end{align}
and \begin{align}
\label{eq2:mainthm}
E(R) &\leq \min_{1\leq j\leq K} B_j\left(1-\frac{R}{C_j}\right), \quad \forall R<\upC.
\end{align}
\end{theorem}
Since the reliability function yields bounds on the capacity of the DM-BC, we immediately  obtain the following.
\begin{corollary} \label{cor:cap_bc} Under the condition $B_{\mathrm{max}}<\infty$, the capacity of the DM-BC with VLFT, namely $C_{\mathrm{BC}\mbox{-}\mathrm{VLFT}} $, satisfies
%\red{not clear why we need $0<B_{\mathrm{min}}$ since this condition is not in the prev theorem: I was wrong, it is a redundant condition} 
\begin{align}
C \leq C_{\mathrm{BC}\mbox{-}\mathrm{VLFT}} \leq \upC. 
\end{align}
\end{corollary}
Although there is, in general, a gap between the upper and lower bounds on the reliability function  (and capacity) provided in Theorem \ref{mainthm} (and Corollary~\ref{cor:cap_bc}), under some conditions on the DM-BC, the reliability function  (and capacity) is known exactly.
\begin{theorem}
\label{stochasticDM}
For a less capable DM-BC   with VLFT such that $B_{\mathrm{max}} <\infty$, %the following holds
\begin{align}
B \left(1-\frac{R}{C_1} \right)\leq E(R) \leq B_1 \left(1-\frac{R}{C_1}\right), \quad \forall R<C_1. \label{eqn:less_cap}
\end{align}
Furthermore, if the DM-BC   with VLFT is stochastically degraded (or physically degraded), %we obtain
\begin{align}
E(R)=B_1 \left(1-\frac{R}{C_1}\right),\quad \forall R<C_1. \label{eqn:stoc_deg}
\end{align}
\end{theorem}
\begin{corollary}  \label{cor:cap} Under the condition $B_{\mathrm{max}}<\infty$, the capacity of any less capable DM-BC with VLFT 
%\red{not clear why we need $0<B_{\mathrm{min}}$ since this condition is not in the prev theorem: I was wrong, it is a redundant condition} 
\begin{align}
  C_{\mathrm{BC}\mbox{-}\mathrm{VLFT}} = C=C_1= \upC.  \label{eqn:cap_less_cap}
\end{align}
\end{corollary}
\begin{IEEEproof}[Proof of Theorem \ref{stochasticDM} and Corollary \ref{cor:cap}]
For any less capable DM-BC we have $I(X;Y_1) \leq I(X;Y_j)$  for all $P_X$ and for all $ j=2,3,\ldots,K$. 
%\red{why do we need the $\tilY_1$ here for less capable? less capable doesn't involve any new rv $\tilY$} and for all $ j=2,3,\ldots,K: This is a typo, I removed it, sorry$.
Hence, 
\begin{align}
C&=\max_{P_X} \min_{1\leq j\leq K} I(X;Y_j),\\
&=\max_{P_X} I(X;Y_1)=C_1.
\end{align} 
Plugging this into \eqref{eq1:mainthm} establishes the lower bound in~\eqref{eqn:less_cap}.   For less capable DM-BCs,  we also have $C_1=\max_{P_X} I(X;Y_1) \leq C_j=\max_{P_X} I(X;Y_j)$ for all $ j=2,3,\ldots,K$, 
hence
\begin{align}
\upC&:=\min_{1\leq j\leq K} \max_{P_X} I(X;Y_j)\\
&=\max_{P_X} I(X;Y_1)=C_1. \label{eqn:cC1}
\end{align}
As a result, for less capable DM-BCs,  the capacity is $C=C_1=\upC$, establishing \eqref{eqn:cap_less_cap}.  
%\red{did you prove the upper bound in \eqref{eqn:less_cap}? i think it's missing. i can't verify the upper bound. Perhaps I'm stupid. pls check}
Moreover, from~\eqref{eq2:mainthm} in Theorem~\ref{mainthm}, for all $R<\upC=C_1$ (cf.~Eqn.~\eqref{eqn:cC1}), 
\begin{align}
E(R)  \leq \min_{1\leq j\leq K} B_j\left(1-\frac{R}{C_j}\right) 
 \leq B_1\left(1-\frac{R}{C_1}\right).
\end{align}
This establishes the upper bound in~\eqref{eqn:less_cap}. 

For   stochastically degraded DM-BCs, there exists a random variable  $\tilde{Y}_1$ such that $X-Y_j-\tilde{Y}_1$ for all $ j=1,2,\ldots,K$ and $P_{\tilde{Y}_1|X} = P_{Y_1|X}$.  Therefore, we have
\begin{align}
D(P_{Y_1|X}(\cdot|x)\|P_{Y_1|X}(\cdot|x'))=D(P_{\tilde{Y}_1|X}(\cdot|x)\|P_{\tilde{Y}_1|X}(\cdot|x')).
\end{align}
%Now, by using the log-sum inequality, it is easy to show that for any $j\in \{2,3,..,K\}$
%\begin{align}
%\label{eq472016}
%D(P_{Y_j|X}(\cdot|x)\|P_{Y_j|X}(\cdot|x')) \geq  D(P_{\tilde{Y}_1|X}(\cdot|x)\|P_{\tilde{Y}_1|X}(\cdot|x')), \forall x, x' \in \calX.
%\end{align}
Observe that for any $x,x' \in \calX$ and $j\in \{2,3,\ldots,K\}$, we also have
\begin{align}
D(P_{Y_1|X}(\cdot|x)\|P_{Y_1|X}(\cdot|x'))&=\sum_{y_1} P_{Y_1|X}(y_1|x)\ln \frac{P_{Y_1|X}(y_1|x)}{P_{Y_1|X}(y_1|x')}\\
&=\sum_{y_1} \sum_{y_j} P_{\tilde{Y}_1 Y_j|X}(y_1 y_j|x)\ln \frac{\sum_{y_j} P_{\tilde{Y}_1 Y_j|X}(y_1 y_j|x)}{\sum_{y_j} P_{\tilde{Y}_1 Y_j|X}(y_1 y_j|x')}\\
&=\sum_{y_1} \sum_{y_j} P_{Y_j|X}(y_j|x)P_{\tilde{Y}_1|Y_j}(y_1|y_j) \ln \frac{\sum_{y_j} P_{Y_j|X}(y_j|x)P_{\tilde{Y}_1|Y_j}(y_1|y_j)}{\sum_{y_j} P_{Y_j|X}(y_j|x')P_{\tilde{Y}_1|Y_j}(y_1|y_j)} \label{eqn:follow_mc}\\
&\le \sum_{y_1} \sum_{y_j} P_{Y_j|X}(y_j|x) P_{\tilde{Y}_1|Y_j}(y_1|y_j) \ln \frac{P_{Y_j|X}(y_j|x)}{P_{Y_j|X}(y_j|x')} \label{eqn:log_sum}\\
&=\sum_{y_j}  P_{Y_j|X}(y_j|x) \ln \frac{P_{Y_j|X}(y_j|x)}{P_{Y_j|X}(y_j|x')} \left(\sum_{y_1} P_{\tilde{Y}_1|Y_j}(y_1|y_j)\right)\\
&=D(P_{Y_j|X}(\cdot|x)\|P_{Y_j|X}(\cdot|x')).
\end{align} Here, \eqref{eqn:follow_mc} follows from the Markov chains $X-Y_j-\tilde{Y}_1 $ for $j=1,2,\ldots,K$ and  \eqref{eqn:log_sum} follows from the log-sum inequality. 

It follows that
\begin{align}
B&=\max_{x,x'\in \calX} \min_{1\leq j\leq K} D(P_{Y_j|X}(\cdot|x)\|P_{Y_j|X}(\cdot|x')) \\
&=\max_{x,x' \in \calX} D(P_{Y_1|X}(\cdot|x)\|P_{Y_1|X}(\cdot|x')) =B_1,
\end{align}
and hence \eqref{eqn:stoc_deg} is established.
\end{IEEEproof} 

A few remarks concerning Theorem \ref{thm:main_res} are in order.
\begin{itemize}
\item There is a gap between the lower and upper bounds for the general DM-BC. One reason that pertains to  the achievability part is because each decoder $j \in \{1,2,\ldots, K\}$, at time $n$, only has its own sequence $Y_j^n$. Thus, it is difficult to establish an appropriate  hypothesis test within the coding scheme by Yamamoto-Itoh~\cite{YamamotoItoh1979} such that this hypothesis test works for any possible realization of the other random variables $\{Y_i^n:i \neq j\}$.
\item For the converse, if we use the same hypothesis test  for single-user channels with VLFT as in Berlin {\em et al.}'s work~\cite{Berlin2009a}, it is challenging to obtain a useful result. The hypothesis test  in~\cite[Prop.~1]{Berlin2009a} involves the  sufficient statistic   $V_n:=\ln {P_\rmA(Y_1^n)}-\ln{P_\rmN(Y_1^n)}$. Because $X_k$ depends on $(W, Y_1^{k-1},\ldots, Y_K^{k-1})$ for each $k\in\bbN$ (cf.~Eqn.~\eqref{eqn:enc}),    we cannot simply append $(Y_2^n, \ldots,Y_K^n)$ to $Y_1^n$ in~the expression for $V_n$ and still obtain the desired upper bound as in~\cite[Prop.~1]{Berlin2009a}. %~\cite[Eqn.~(9)]{Berlin2009a} for the relevant sufficient statistic in the  hypothesis test $V_n:=\ln {P_\rmA(Y_1^n)}-\ln{P_\rmN(Y_1^n)}$. %\red{should this be $V_n:=\ln {P_\rmA(Y_1^n)}-\ln{P_\rmN(Y_1^n)}$?}. 
\item Moreover, if we directly adapt  the  key ideas in Burnashev's converse proof for sequential hypothesis testing in~\cite[Lemmas~3 and~4]{Burnashev80}, we will only obtain the following  almost sure bound for each $j\in\{1,\ldots, K\}$:
\begin{align}
&\bbE\left[\calH(W|Y_j^n)-\calH(W|Y_j^{n+1})|Y_j^n \right] \nn\\
\label{eq47remark}
&\leq \max_{w, w' \in \calW}\sup_n \sup_{ y_j^{n-1}} D \big(P_{Y_{j,n}|Y_j^{n-1},W}(\cdot|y_j^{n-1},w)\,\big\|\,P_{Y_{j,n}|Y_j^{n-1},W}(\cdot|y_j^{n-1},w') \big).
\end{align}
This is then insufficient to establish our converse. 
%Hence, we need to adapt Burnashev's ideas in \cite{Burnashev80} and also Burnashev's work on sequential hypothesis testing in \cite{Burnashev1976} for our setting. 
\item Our Lemma~\ref{lem5new} is stronger than the corresponding one to prove the converse of~\eqref{eqn:burn} in Burnashev~\cite[Lemma 3]{Burnashev1976} since we do not need to assume that the conditional entropies $\calH(W|Y_j^n)$ for $j=1,2,\ldots,K$ are bounded. Consequently, the construction of submartingales in the proof of Lemma~\ref{conversethm}  (in the converse proof in Section \ref{sec:conveseproof}) is much simpler.
\item We have a tight reliability function result for stochastically degraded DM-BCs in \eqref{eqn:stoc_deg}. Usually, orderings of the channels (less/more capable, less noisy, stochastically and physically degraded) are used to obtain tight capacity or capacity region results for DM-BCs~\cite[Secs.~3.4~\&~3.6]{elgamal}. Here, in contrast,  we use the orderings to establish   a tight reliability function result. 
\end{itemize} %}

\section{Achievability Proof of Theorem \ref{thm:main_res} }\label{achproof}
In this section, we provide the achievability proof of Theorem \ref{thm:main_res}. We start with  a preliminary lemma.
\begin{lemma}[Expectation of the Maximum of Random Variables]\label{lem1b}
Let $\{(X_{1L},X_{2L},\ldots,X_{KL})\}_{L\ge 1}$ be $K$ sequences of random variables satisfying %the following properties for some  {constants} $A\geq 0, D\geq 0$ and $B_1,B_2,B_3 \in \bbR$:
\begin{align}
\bbE [X_{jL} ]&= L +o(1), \quad\mbox{and}\\
\var(X_{jL}) &= o(1),\quad j=1,2,\ldots,K,
\end{align}
as $L\to\infty$. Then, as $L\to\infty$, we have
\begin{align}
\bbE(\max\{X_{1L},X_{2L},\ldots, X_{KL}\}) = L+O(\sqrt{L}). \label{eqn:exp_max}
\end{align}
\end{lemma}
\begin{IEEEproof} The proof can be found in  Appendix \ref{app:exp_rvs}. \end{IEEEproof}
The achievability part of Theorem \ref{thm:main_res} can be  stated succinctly as follows.
\begin{lemma}
\label{yamamotolemma}
If $B_{\mathrm{max}} <\infty$,  
\begin{align}
\label{ach:errexp}
E(R) \geq B \left(1-\frac{R}{C}\right),\quad \forall R< C.
\end{align}
\end{lemma}
\begin{IEEEproof} The achievability proof is an extension of   Yamamoto-Itoh's variable-length coding scheme~\cite{YamamotoItoh1979} for the DMC with noiseless variable-length feedback. However, we devise  some additional and crucial ingredients to account for the presence of multiple channel outputs and multiple decoded messages. In the coding scheme, the encoder decides whether or not to stop the transmission. We   show that for all $L\in\bbN$ there exists an $(\lceil e^{RL}\rceil, L+O(\sqrt{L}))$-VLFT code with achievable exponent $B \left(1-R/C\right)$. 

Choose  $P_X^*:=\argmax_{P_X} \min_{1\leq j\leq K} I(X;Y_j)$ and $x_{\rmc}, x_{\rme} \in \calX$ such that
\begin{align}
(x_{\rmc},x_{\rme}):=\argmax_{(x,x')\in \calX} \min_{1\leq j\leq K} D\big(P_{Y_j|X}(\cdot|x)\|P_{Y_j|X}(\cdot|x') \big).
\end{align}
%This means that $\min_{1\leq j\leq K} I(X;Y_j)=C$ under the input distribution $P_X^*$. Besides, 
Since we assume that $B_{\mathrm{max}} <\infty$, we have $P_{Y_j|X}(y|x)>0$ for all $y \in \calY_j, x\in \calX$ for all $j=1,2,\ldots,K$. Fix a non-negative number $R$ satisfying $0\leq R<C$.

We design a code for each block of $L$ transmissions as per the Yamamoto-Itoh coding scheme with rate $R$~\cite{YamamotoItoh1979}. Let this code length $L$ be divided into two parts, $\gamma L$ for the message mode and $(1-\gamma)L$ for the control mode.  In the message mode, one of $M=\lceil e^{LR}\rceil$ messages is \emph{transmitted by a random coding scheme} with block-length $\gamma L$~\cite{Gallager1965a}, and in the control mode a pair of control signals $(\rmc, \rme)$ is transmitted by another block code with length $(1-\gamma)L$. The control signal $\rmc$ is only sent   when all the $K$ receivers correctly decode  the transmitted message in the message mode. 

Now, the variable-length coding scheme for the DM-BC with a common message  is created by repeating the length-$L$ transmission at times $n\in \{\mu L: \mu=1,2,3,\ldots\}$ and using the same decoding algorithm  as in~\cite{YamamotoItoh1979} at all the decoders. The decoder $j \in \{1,2,\ldots, K\}$ defines a stopping time $\tau_j$ as follows:
\begin{enumerate}
\item If $n\in \{\mu L:\mu=2,3,4,\ldots\}$, we define
\begin{align}
\bone\{\tau_j=n\}& = \prod_{t=1}^{\mu-1} \bone\left\{g_n^{(j)}\left(Y_{j,(t-1)L+\gamma L+1}^{(t-1)L+L}\right)=\rme\right\}  \bone\left\{g_n^{(j)}\left(Y_{j,(l-1)L+\gamma L+1}^{n}\right)=\rmc\right\};
\end{align}
\item If $n = L$, we define 
\begin{align}
\bone\{\tau_j=n\}=\bone \left\{g_n^{(j)}\left(Y_{j,\gamma L+1}^{L}\right)=\rmc\right\};
\end{align}
\item Otherwise,
\begin{align}
\bone\{\tau_j=n\}=\bone\{\emptyset\}.
\end{align}
\end{enumerate}
%in the equation~\eqref{def_tau} at the top of this page.
%\begin{figure*}[!t]
% ensure that we have normalsize text
%\normalsize
% Store the current equation number.
%\setcounter{MYtempeqncnt}{\value{27}}
% Set the equation number to one less than the one
% desired for the first equation here.
% The value here will have to changed if equations
% are added or removed prior to the place these
% equations are referenced in the main text.
%\setcounter{equation}{27}
%\begin{align}
%\label{def_tau}
%&1\{\tau_j=n\}=\begin{cases}\prod_{t=1}^{l-1} 1\left\{g_n^{(j)}\left(Y_{j,(t-1)N+1}^{(t-1)N+N}\right)=e\right\}1\left\{g_n^{(j)}\left(Y_{j,(l-1)N+1}^{n}\right)=c\right\},&\mbox{if}\quad n\in \{lN:l=2,3,4,\ldots\}\\ 1\left\{g_n^{(j)}\left(Y_{j}^{N}\right)=c\right\},&\mbox{if}\quad n=N\\
%\\0,&\mbox{otherwise}  \end{cases}\\
%\label{def_P_E}
%&\rvP_{\rmE}=\rvP\left(\{\emph{There exists a decoder that decodes a wrong message at data mode but decodes the message $c$ at control mode } \}\right),\\
%\label{def_P_X}
%&\rvP_{\rmX}=\rvP\left(\{\emph{There exist a decoder that decodes the message $e$ at the control mode}\}\right).
%\end{align}
%% Restore the current equation number.
%%\setcounter{equation}{\value{MYtempeqncnt}}
%% IEEE uses as a separator
%\hrulefill
%% The spacer can be tweaked to stop underfull vboxes.
%\vspace*{4pt}
%\end{figure*}
%--------------------------------------------------
In addition, the estimated message at the stopping time $\tau_j$ has the following form:
\begin{align}
\hat{W}_j:=g_{\tau_j}^{(j)}\left(Y_{j,\tau_j-L}^{\tau_j- (1-\gamma)L  }\right),\quad j=1,2,\ldots, K.
\end{align}
Since $\calY_j$ for $ j\in\{1,2,\ldots,K\}$ is finite, for each fixed $n \in \bbZ_{+}$ all the decoding regions at each decoder $j$ are finite sets, which are Borel sets in $\bbR^n$. Combining this fact with the definition of $\tau_j$, we have $\bone\{\tau_j=n\}\in \sigma(Y_j^n)$ for all $ n\in\bbN$. Let 
\begin{align}
q_L^{(j)}:=\bbP\left(g_n^{(j)}(Y_{j,\gamma L+1}^L)=\rme\right), \quad j=1,2,\ldots,K.
\end{align}
By the proposed  transmission method, given $W=w\in \calW$ we have that $Y_{j,(t-1)L+1}^{(t-1)L+L}$ for $t\in\bbN$ are independent random vectors. Since  the messages in $ \calW$ are equiprobable, we obtain
\begin{align}
\bbP(\tau_j=n)=\begin{cases} \big[q_L^{(j)}\big]^{l-1} \big[1-q_L^{(j)}\big],&\mbox{if}\quad n\in \{\mu L: \mu=1,2,3,\ldots\}\\ 0,&\mbox{otherwise}\end{cases}.
\end{align}
Hence, we have
\begin{align}
\sum_{n=0}^{\infty} \bbP(\tau_j=n) = \sum_{\mu=1}^{\infty} \big[q_L^{(j)}\big]^{\mu-1} \big[1-q_L^{(j)}\big]=1.
\end{align}
Thus,   $\tau_j$ is a stopping time with respect to $\{\sigma(Y_j^n)\}_{n=0}^{\infty}$.

Now, since we use the same decoding algorithm as~\cite{YamamotoItoh1979} for each repeated transmission block of length $L$   at each decoder $j$, it is easy to see that the error probability for the $j$-th decoder $\rvP_{\rmE}^{(j)}:=\bbP(\hat{W}_j \neq W)$ and $q_L^{(j)}$ can be written as follows~\cite{YamamotoItoh1979}:
\begin{align}
\label{eq41new}
\rvP_{\rmE}^{(j)}&= \rvP_{1\rme}^{(j)}\rvP_{2\rme\rmc}^{(j)}, \\
q_L^{(j)}&=\rvP_{1\rme}^{(j)}(1-\rvP_{2\rme\rmc}^{(j)})+(1-\rvP_{1\rme}^{(j)})\rvP_{2\rmc\rme}^{(j)}.
\end{align}
Here, $\rvP_{1\rme}^{(j)}$, $\rvP_{2\rme\rmc}^{(j)}$, and $\rvP_{2\rmc\rme}^{(j)}$ respectively denote  the error probability of decoder $j$ in the message mode, the probability that the  message $\rme$ is sent at the control mode but the decoder $j$ decodes  the message $\rmc$, the probability that $\rmc$ is sent at the control mode but the decoder $j$ decodes  $\rme$~\cite[pp.~730]{YamamotoItoh1979}.

Since $q_L^{(j)}$ is the same for all repeated transmissions, each of blocklength $L$, we have for all $j=1,2,\ldots,K$,
\begin{align}
\bbE(\tau_j)&= \sum_{n=0}^{\infty} n \bbP(\tau_j=n)\\
&=\sum_{\mu=1}^{\infty} \mu L \big[q_L^{(j)}\big]^{\mu-1} \big[1-q_L^{(j)}\big]\\ 
\label{eq59newest}
&=\frac{L}{1-q_L^{(j)}}.
\end{align}  In addition, we also have
\begin{align}
\label{eq60newest}
\var(\tau_j)=\frac{L^2 q_L^{(j)}}{\big[1-q_L^{(j)}\big]^2}.
\end{align}

Let $l:=(1-\gamma)L$. We assign length-$l$    codewords $X_{\rmc}^l=(x_{\rmc},x_{\rmc},\ldots,x_{\rmc}) \in\calX^l$ and ${X}_{\rme}^l=(x_{\rme},x_{\rme},\ldots,x_{\rme})\in\calX^l$ % \red{how long are these codewords: ANSWER: The length of all these sequences are equal to the size of control mode, i.e. $(1-\gamma)N$} 
to control the signals $\rmc$ and $\rme$ respectively. Decoding of the control signal is done as follows. Choose an arbitrarily small  $\delta>0$. Let us say the number of output symbols $y \in \calY_j$ contained in the received sequence $Y_j^l=y_j^l$	equals to $l_y\in\{1,\ldots,l\}$.  We suppress the dependence of $l_y$ on $j$ for notational convenience. If every $l_y$ satisfies the  typicality condition 
\begin{align}
(1-\delta) P_{Y_j|X}(y|x_{\rmc}) \leq  \frac{l_y }{l} \leq (1+\delta)  P_{Y_j|X}(y|x_{\rmc}),
\end{align} then $y_j^l$ %\red{why is the length $n$? shouldn't it be $(1-\gamma)L$?}
 is decoded to $\rmc$, otherwise to $\rme$.  Then, {defining $F(\cdot)$ to be the random coding error exponent for DMCs~\cite{Gallager1965a} and $R_{L\gamma}:=R/\gamma< \min_{1\leq j\leq K} I(X;Y_j)=C$ (since $X\sim P_X^*$),  it follows from~\cite{YamamotoItoh1979} that}
\begin{align}
\label{keynew}
\rvP_{1\rme}^{(j)} &\dotleq \exp\left[-\gamma L F(R_{L\gamma})\right], \\
\rvP_{2\rmc\rme}^{(j)} &\dotleq \exp\left[-(1-\gamma) L (f_j(\delta)-o(1))\right],
\label{keynew1980}
\end{align} %\red{removed $J_j$ since it's a const}
where $f_j(\delta)>0$ for any $\delta>0$. In  \eqref{keynew} and \eqref{keynew1980} we used the usual notation $a_L\dotleq b_L$ to mean that $\limsup_{L\to\infty}\frac{1}{L}\log\frac{a_L}{b_L}\le 0$. 
Also, by Stein's lemma, 
\begin{align}
\label{stein}
\lim_{L\to \infty} -\frac{\ln \rvP_{2\rme\rmc}^{(j)}}{(1-\gamma) L}= D \big(P_{Y_j|X}(\cdot|x_{\rmc})\|P_{Y_j|X}(\cdot|x_{\rme})\big).
\end{align}
%where $F(R_{L\gamma})>0, J_j>0$ if $0\leq R_{L\gamma}=R/\gamma< \min_{1\leq j\leq K} I(X;Y_j)=C$. %Here, $F(R_{L\gamma})$ is the regular (usual) random coding exponent for broadcast channel with common messages (no feedback).
%\red{this is not the definition of $C$; $C$ is $\max_{P_X}\min_{1\leq j\leq K} I(X;Y_j)$ right?: This is the definition of C since we choose distribution P_X in right before (55).} 
%\red{why do we need this since you defined $\delta$?: It is redundant from previous version, I forgot to get rid of it, sorry} 
%Note that $f_j(\delta)>0$ since $\delta>0$ for all $j=1,2,\ldots,K$. 
%\red{i also think we don't need this statment: Yes, it only affirm that $f_j(\delta)>0$ if $delta>0$. You can remove it if you want}

Moreover from~\eqref{eq41new} and~\eqref{keynew}--\eqref{keynew1980} we have
\begin{align}
\label{important2}
q_L^{(j)} \dotleq  \exp(-L c^{(j)}  ),\quad j=1,2,\ldots,K
\end{align} for some exponent $c^{(j)}>0$. %\red{removed $a_X^{(j)} $ as it's a const}

Consequently, from~\eqref{eq59newest},~\eqref{eq60newest}, and~\eqref{important2} %\ red{I think this label~\eqref{important2} is incorrect. It is correct since if we replace ~\eqref{important2} to ~\eqref{eq59newest},~\eqref{eq60newest} we obtain the following} 
we obtain for all $j$ that
\begin{align}
\label{veryimporant11}
\bbE(\tau_j)&=L+o(1),\\
\label{veryimporant12}
\var(\tau_j)&=o(1).
\end{align}
From~\eqref{veryimporant11},~\eqref{veryimporant12}, and Lemma~\ref{lem1b} we obtain that
\begin{align}
\label{keypointnew}
\bbE(\tau) = L+O(\sqrt{L}).
\end{align}
Now, since for each $j = 1,2,\ldots, K$,  $\rvP_{\rmE}^{(j)}$ is kept the same for all repeated transmission blocks of length $L$, we have
\begin{align}
\label{eq113new2016}
\rvP_{\rme}(R,L+O(\sqrt{L}))\leq \sum_{j=1}^K \rvP_{\rmE}^{(j)}.
\end{align} 
%Note that
%\begin{align}
%\liminf_{\bbE(\tau)\to \infty}\frac{\ln 2^{NR}}{\bbE(\tau)}=\liminf_{N\to \infty}\frac{\ln 2^{NR}}{\gamma (N+O(\sqrt{N}))}=R.
%\end{align} 
Moreover, it is easy to see from~\eqref{eq41new},~\eqref{keynew}--\eqref{keynew1980}, and~\eqref{keypointnew} that $\rvP_{\rmE}^{(j)}\to 0$ for all $j=1,2,\ldots, K$ as $L \to \infty$ if $0\leq R_{L\gamma}=R/\gamma<   C$ and $0\leq \gamma<1$. Combining these requirements and~\eqref{eq113new2016}, we have $\rvP_{\rme}(R,L+O(\sqrt{L})) \to 0$ as $L\to \infty$ if we choose $1>\gamma >R/C$.  Now, since $\gamma> R/C$, a feasible value of $\gamma$ that we can choose is 
\begin{align}
\label{gammakey}
\gamma=\frac{R}{C-\eps},
\end{align}
where $\eps>0$ is chosen small enough so that $\gamma$ remains smaller than $1$. It follows that for any $R\in [0,C)$, we have
\begin{align}
\liminf_{L\to \infty} -\frac{\ln\rvP_{\rme}(R,L+O(\sqrt{L}))}{L+O(\sqrt{L})} &\geq \liminf_{L \to \infty}- \frac{\ln\big(\sum_{j=1}^K \rvP_{\rmE}^{(j)}   \big)  }{L+O(\sqrt{L})}\\
&\geq \liminf_{L \to \infty} \left\{\min_{1\le j\le K}- \frac{\ln( K \rvP_{\rmE}^{(j)})}{L+O(\sqrt{L})} \right\}\\
&=  \min_{1\leq j\leq K} \left\{ \liminf_{L \to \infty} - \frac{\ln \rvP_{\rmE}^{(j)}}{L+O(\sqrt{L})}\right\} \label{eqn:liminf_min}\\
&\geq \min_{1\leq j\leq K} \left\{\lim_{L\to \infty} - \frac{\ln \rvP_{2\rme\rmc}^{(j)}}{L}\right\}\label{eqn:yi}\\
&=\min_{1\leq j\leq K} D(P_{Y_j|X}(\cdot|x_{\rmc})\|P_{Y_j|X}(\cdot|x_{\rme}))\left(1-\frac{R}{C-\eps}\right)\label{eqn:yi2} \\
\label{eqkeylan}
& = B\left(1-\frac{R}{C-\eps}\right),
\end{align}
where \eqref{eqn:liminf_min} follows from the facts that $K$ is a  constant and that $\liminf_{L\to\infty} \min_j \{a_{jL}\}=\min_j \liminf_{L\to\infty} \{a_{jL}\}$ for any family of sequences $\{a_{jL}\}$; \eqref{eqn:yi} follows from~\eqref{eq41new}; and~\eqref{eqn:yi2} follows from~\eqref{stein} and~\eqref{gammakey}.

This means that $(R, B(1-R/(C-\eps)))$ is an achievable rate-exponent pair for any $0\leq R<C$. By the arbitrariness of $\eps>0$, we obtain
\begin{align}
\label{eqkeylan2}
E(R)\geq B \left(1-\frac{R}{C}\right).
\end{align}

Finally, for any $N\in \bbR_+$ choose  $L=\lfloor N-O(\sqrt{N}) \rfloor$ such that $L+O(\sqrt{L})\leq N$. By using the $(\lceil e^{RL}\rceil, L+O(\sqrt{L}))$-VLFT code  constructed above, we conclude that there exists an $(\lceil e^{\lfloor (N-O(\sqrt{N}))R\rfloor }\rceil,N)$-VLFT code such that~\eqref{ach:errexp} holds. 
\end{IEEEproof}
 
 We remark that  for the proof of Lemma~\ref{yamamotolemma}, we extended Yamamoto and Itoh's coding scheme~\cite{YamamotoItoh1979} for the DM-BC with a common message  and VLFT.  In the proof, we supplemented some new elements to the original argument in \cite{YamamotoItoh1979}. These include  defining appropriate stopping times $\{\tau_1,\tau_2,\ldots , \tau_K\}$ and proving that the expectation  of the maximum of these  $K$ stopping times  with expectations and variances respectively bounded by $L+o(1)$  and $o(1)$  is $L+O(\sqrt{L})$ (cf.\ Lemma~\ref{lem1b}).% READ THIS TO MAKE SURE I DIDN'T DISTORT YOUR INTENTION: YES, THE STATEMENT IS THE SAME AS WHAT I WOULD LIKE TO SAY}
 
\section{Converse Proof of Theorem \ref{thm:main_res}}\label{sec:conveseproof}
In this section,  we provide the converse   proof of Theorem \ref{thm:main_res}. We start with  a few preliminary lemmas. At the end of the proof (after the proof of Lemma \ref{conversethm}), we discuss the novelites in our converse proof vis-\`a-vis Burnashev's works in~\cite{Burnashev1976} and~\cite{Burnashev80}.
\begin{lemma}
\label{lem1newest}
Under the condition that $\bbP(\tau <\infty)=1$ (cf.\ Definition \ref{def1}), the following inequalities hold 
\begin{align}
\bbE\left[\calH(W|Y_j^{\tau_j})\right] &\leq h(\rvP_{\rme}(R_N,N))+\rvP_{\rme}(R_N,N)\ln(M-1),
\end{align} for each $1\leq j\leq K$ and $N$ sufficiently large. %\red{do you mean $R$ or $R_N$, $M$ or $M_N$? Is $R$ being achievable necessary?: ANSWER: I meant $R_N$ and $M$ as the Lemma statements}
\end{lemma}
\begin{IEEEproof}
The proof of this Lemma is essentially the same as~\cite[Lemma~1]{Burnashev80}. For completeness and compatibility in the notations, we provide the complete proof in   Appendix~\ref{append}. Note that the error event here is different from~\cite[Lemma~1]{Burnashev80}. It is the union of error events of individual branches of the DM-BC, i.e., $\cup_{j=1}^K \{\hat{W}_j \neq W\}$. 
\end{IEEEproof}
\begin{lemma}
\label{lem3new}
 For any $n\geq 0$ the following inequalities hold almost surely (cf.~Definition~\ref{def:info_q})
\begin{align}
\bbE[\calH(W|Y_j^n)- \calH(W|Y_j^{n+1})|Y_j^n] &\leq C_j, \quad 1\leq j\leq K.
\end{align}
\end{lemma}
\begin{IEEEproof}
Observe that
\begin{align}
\bbE[\calH(W|Y_1^n)- \calH(W|Y_1^{n+1})|Y_1^n]&=\bbE[\calH(W|Y_1^n)- \calH(W|Y_1^{n+1})|Y_1^n]\\
&= \bbE[\calI(W;Y_{1,n+1}|Y_1^n)|Y_1^n]\\
&= \calI(W;Y_{1,n+1}|Y_1^n)\\
&\leq \calI(W,X_{n+1};Y_{1,n+1}|Y_1^n)\\
\label{eq53new}
&\leq \calI(X_{n+1};Y_{1,n+1}|Y_1^n)+ \sum_{x\in \calX} \calI(W;Y_{1,n+1}|X_{n+1}=x,Y_1^n)\bbP(X_{n+1}=x|Y_1^n).
\end{align} 
Now, for any fixed $Y_1^n=y_1^n$, the (random) mutual information in the sum can be expressed as % \red{what's the diff between the first two expressions below concerning $I$ and $\calI$}
\begin{align}
&\calI(W;Y_{1,n+1}|X_{n+1}=x,Y_1^n=y_1^n)\nn\\* 
&=I(W;Y_{1,n+1}|X_{n+1}=x,Y_1^n=y_1^n)\\
&=\sum_{w\in \calW, y\in \calY_1} \bbP(W=w,Y_{1,n+1}=y|X_{n+1}=x,Y_1^n=y_1^n)\nn\\
\label{eq55new}
& \qquad \times \ln \frac{\bbP(W=w,Y_{1,n+1}=y|X_{n+1}=x,Y_1^n=y_1^n)}{\bbP(W=w|X_{n+1}=x,Y_1^n=y_1^n)\bbP(Y_{1,n+1}=y|X_{n+1}=x,Y_1^n=y_1^n)}.
\end{align}
Since $(W,Y_1^n,Y_2^n,\ldots,Y_K^n)-X_{n+1}-(Y_{1,n+1},Y_{2,n+1},\ldots,Y_{K,n+1})$ forms a Markov chain, we obviously  also have the following Markov chain: 
\begin{align}
  (W,Y_1^n)-X_{n+1}-Y_{1,n+1}.\label{eqn:mc1} %,\quad\mbox{and}% \label{eqn:mc1}\\ 
 %& Y_1^n  -X_{n+1}-Y_{1,n+1} .\label{eqn:mc2}
\end{align}
 Hence, we have
\begin{align}
&\bbP(W=w,Y_{1,n+1}=y|X_{n+1}=x,Y_1^n=y_1^n)\\
&=\bbP(W=w|X_{n+1}=x,Y_1^n=y_1^n) \bbP(Y_{1,n+1}=y|X_{n+1}=x,Y_1^n=y_1^n, W=w)\\
&=\bbP(W=w|X_{n+1}=x,Y_1^n=y_1^n) \bbP(Y_{1,n+1}=y|X_{n+1}=x)\\
&=\bbP(W=w|X_{n+1}=x,Y_1^n=y_1^n) \bbP(Y_{1,n+1}=y|X_{n+1}=x,Y_1^n=y_1^n).
\end{align}
From~\eqref{eq55new} we obtain
\begin{align}
\calI(W;Y_{1,n+1}|X_{n+1}=x,Y_1^n=y_1^n)=0,\quad \forall (x,y_1^n) \in \calX \times \calY_1^n.
\end{align}
It follows from~\eqref{eq53new} that
\begin{align}
\bbE[\calH(W|Y_1^n)- \calH(W|Y_1^{n+1})|Y_1^n]&\leq \calI(X_{n+1};Y_{1,n+1}|Y_1^n)\\
&\leq C_1,\quad a.s.
\end{align}
A completely analogous argument goes through to yield the corresponding upper bounds for $j=2,3,\ldots,K$. 
%Similarly, we have
%\begin{align}
%\bbE[\calH(W|Y_j^n)- \calH(W|Y_j^{n+1})|Y_j^n] \leq C_j, \quad a.s. \quad j=2,3,\ldots,K.
%\end{align} 
\end{IEEEproof}
We remark that in the  above proof, we need to use some additional arguments involving the Markov chain  in \eqref{eqn:mc1}   to show that   Lemma~\ref{lem3new} holds in the (general DM-BC)  case where  $X_{n+1}$ is a function of $W$ and {\em all} $Y_j^n$ for $j=1,2,\ldots,K$. In the   DMC, $X_{n+1}$ is  a function of $W$ and  {\em only} $Y_1^n$.% for {\em some} $j\in \{1,2,\ldots,K\}$. 

%\red{PLEASE CHECK CORRECTNESS OF THIS STATEMENT: ANSWER: YES, IT IS CORRECT}
The following lemma is a restatement of \cite[Lemma 7]{Burnashev1976}.
\begin{lemma}
\label{lem3} For arbitrary non-negative numbers $p_l, f_i, \beta_{il}$ where $l=1,2,\ldots,L$ and  $i=1,2,\ldots,N$, we have the following inequality
\begin{align}
\sum_{l=1}^L p_l \ln \left(\frac{\sum_{i=1}^N f_i}{\sum_{i=1}^N \beta_{il}}\right) \leq \max_i \sum_{l=1}^L p_l \ln \frac{f_i}{\beta_{il}}.
\end{align}
\end{lemma}
\begin{lemma} For any $n\geq 0$ the following inequalities hold almost surely (cf.~Definition \ref{def:info_q})
\label{lem5new}
\begin{align}
\label{eq66new}
\bbE[\ln\calH(W|Y_j^n)-\ln \calH(W|Y_j^{n+1})|Y_j^n] &\leq B_j, \quad 1\leq j\leq K.
\end{align} 
\end{lemma}
\begin{IEEEproof}
The proof is based on Burnashev's arguments in~\cite{Burnashev1976} and~\cite{Burnashev80} with some modifications to account for the fact that at each transmission time $n+1$, the transmitted signal $X_{n+1}$ is a function of $W$ and {\em all} $Y_1^n, Y_2^n,\ldots,Y_K^n$. We can assume that $P_{Y_j|X}(y_j|x)>0$ for all $ x \in \calX, y_j \in \calY_j $ and all  $ j=1,2,\ldots,K$, otherwise the inequalities~\eqref{eq66new} trivially hold  since $B_j=\infty$. For each $i=1,2,\ldots,M$ and $y\in\calY_1$, define
\begin{align}
\label{eq70}
p_i&:=\bbP(W=i|Y_1^n),  \\
p_i(y)&:=\bbP(W=i|Y_1^n, Y_{1,n+1}=y),\label{eqn:piy} \\
p(y|W=i)&:=\bbP(Y_{1,n+1}=y|Y_1^n, W=i),\\
p(y|W\neq i)&:=\bbP(Y_{1,n+1}=y|Y_1^n, W\neq i),\\
\label{eq73}
p(y)&:=\bbP(Y_{1,n+1}=y|Y_1^n).
\end{align}
We may assume without loss of generality that $p_i \neq 1$ for all $ i \in\calW=\{1,\ldots,M\}$. Otherwise, again the inequalities in~\eqref{eq66new} trivially hold. Using Lemma~\ref{lem3} and the definitions in~\eqref{eq70}--\eqref{eq73} we have
\begin{align}
\label{eq75new}
\bbE\left[\ln \calH(W|Y_1^n)-\ln \calH(W|Y_1^{n+1})\,\big|\,Y_1^n\right]&=\sum_{y \in \calY_1} p(y)\ln\left[\frac{-\sum_{i=1}^M p_i \ln p_i}{-\sum_{i=1}^M p_i(y) \ln p_i(y) }\right]\\
\label{eq76new}
&\leq \max_{i} \left\{\sum_{y \in \calY_1} p(y)\ln\left[\frac{- p_i \ln p_i}{-p_i(y) \ln p_i(y) }\right] \right\}  
\end{align}
Define % \red{I REMOVED TWO MINUSES. PLEASE CHECK THE DEF OF $F_i$ below}
\begin{equation}
F_i:=\sum_{y \in \calY_1} p(y)\ln\left[\frac{ - p_i \ln p_i}{- p_i(y) \ln p_i(y) }\right] 
\end{equation}
It is easy to see that
\begin{align}
\label{eq77new}
p(y)&=p_i p(y|W=i)+(1-p_i) p(y|W\neq i),\\
\label{eq78new}
p_i(y)&=\frac{p_i p(y|W=i)}{p(y)},
\end{align}
and
\begin{align}
p(y|W=i)&=\bbP(Y_{1,n+1}=y|Y_1^n, W=i)\\
&=\sum_{x \in \calX} \bbP(X_{n+1}=x|W=i,Y_1^n) \bbP(Y_{1,n+1}=y|X_{n+1}=x, W=i,Y_1^n)\\
&=\sum_{x \in \calX} \bbP(X_{n+1}=x|W=i,Y_1^n) \bbP(Y_{1,n+1}=y|X_{n+1}=x) \label{eqn:use_mc}\\
&=:\sum_{x \in \calX} \alpha_{ix} P_{Y_1|X}(y|x). \label{eqn:inv}
\end{align} Here, \eqref{eqn:use_mc} follows from the Markov chain $(W,X_1^n,X_2^n,\ldots,X_K^n)-X_{n+1}-(Y_{1,n+1},Y_{2,n+1},\ldots,Y_{K,n+1})$ and  \eqref{eqn:inv} follows from the invariance (stationarity) of the distribution $\bbP(Y_{1,n+1}=y|X_{n+1}=x)$ in $n$, which is derived from the invariance of the distribution $\bbP(Y_{1,n+1}=y_1,Y_{2,n+1}=y_2,\ldots,Y_{K,n+1}=y_K|X_{n+1}=x)$ in $n$.
Similarly, we have
\begin{align}
p(y|W\neq i)&=\bbP(Y_{1,n+1}=y|Y_1^n, W\neq i)\\
&=\sum_{x \in \calX} \bbP(X_{n+1}=x|W\neq i,Y_1^n) \bbP(Y_{1,n+1}=y|X_{n+1}=x, W\neq i,Y_1^n)\\
&=\sum_{x \in \calX} \bbP(X_{n+1}=x|W\neq i,Y_1^n) \bbP(Y_{1,n+1}=y|X_{n+1}=x)\\
& =:\sum_{x \in \calX} \beta_{ix} P_{Y_1|X}(y|x).
\end{align}
It is easy to see that for each fixed message $i \in \calW = \{1,\ldots, M\}$ we have
\begin{align}
\label{eqkey}
\sum_{x \in \calX} \alpha_{ix}=\sum_{x\in \calX} \beta_{ix}=1, \quad \alpha_{ix}\geq 0, \beta_{ix} \geq 0.
\end{align}

Observe that $F_i$ is a function of variables $p_i, \{\alpha_{ix}\}$ and $\{\beta_{ix}\}$.  For the purpose of finding an upper bound on $\max_i \{F_i\}$ in~\eqref{eq76new}, we can consider only the constraints in~\eqref{eqkey} and find the maximization of $F_i$ over this convex set since other constraints that define the feasible set will only make $F_i$ smaller. With this consideration, let us consider find the maximization of $F_i$ over $\{\beta_{ix}\}$ with the assumption that $\sum_{x\in \calX} \beta_{ix}=1$ and $\beta_{ix} \geq 0$. Fix an arbitrary $x' \in \calX$, then we have $\beta_{ix'}=1-\sum_{x\in \calX\setminus\{x'\}}\beta_{ix}$. We readily obtain that the derivatives of $F_i$ for any $x\in \calX\setminus \{x'\}$ are
\begin{align}
\label{eq88}
\frac{\rmd^2 F_i}{\rmd\beta_{ix}^2}&=\frac{\partial^2 F_i}{\partial \beta_{ix}^2}+ \frac{\partial^2 F_i}{\partial \beta_{ix'}^2}-2 \frac{\partial^2 F_i}{\partial \beta_{ix} \partial \beta_{ix'}},\\
\frac{\partial^2 F_i}{\partial \beta_{ix} \partial \beta_{ix'}}&=(1-p_i)^2 \sum_{y \in \calY_1} \frac{\partial^2 F_i}{\partial p(y)^2} P_{Y_1|X}(y|x) P_{Y_1|X}(y|x'),\\
\label{eq90}
\frac{\partial^2 F_i}{\partial p(y)^2}&=\frac{1}{p(y)}\left[1-\left(\ln \frac{p(y)}{p_i p(y|W=i)}\right)^{-1}+\left(\ln \frac{p(y)}{p_i p(y|W=i)}\right)^{-2}\right] > 0.
\end{align}
Hence, from~\eqref{eq88} to~\eqref{eq90} we obtain
\begin{align}
\label{eq91new}
\frac{\rmd^2 F_i}{\rmd\beta_{ix}^2} =(1-p_i)^2 \sum_{y \in \calY_1} \frac{\partial^2 F_i}{\partial p(y)^2} \left(P_{Y_1|X}(y|x)-P_{Y_1|X}(y|x')\right)^2\geq 0, 
\end{align}
for any $x \in \calX\setminus\{x'\}$.

If for all $x\in \calX\setminus\{x'\}$ we have $D(P_{Y_1|X}(\cdot|x)\|P_{Y_1|X}(\cdot|x'))=0$, it follows that 
%\red{For any $x\in \calX\setminus\{x'\}$ such that  $D(P_{Y_1|X}(\cdot|x)\|P_{Y_1|X}(\cdot|x'))=0$: I changed above since the following does not hold with this assumption. I have a wrong proof in the past for this part} 
\begin{align}
p(y|W=i)&=\sum_{x\in \calX} \alpha_{ix} P_{Y_1|X}(y|x)\\
&=\sum_{x\in \calX} \alpha_{ix} P_{Y_1|X}(y|x')\\
&=\sum_{x\in \calX-\{x'\}} \alpha_{ix} P_{Y_1|X}(y|x')+\alpha_{ix'} P_{Y_1|X}(y|x') \\
&=(1-\alpha_{ix'})  P_{Y_1|X}(y|x')+ \alpha_{ix'} P_{Y_1|X}(y|x') \\
&=(1-\alpha_{ix'}) P_{Y_1|X}(y|x)+ \alpha_{ix'} P_{Y_1|X}(y|x) \\
&=P_{Y_1|X}(y|x), 
\end{align}
for any $i\in\calW$ and $y\in \calY_1$. 
In combination with the fact that the message is uniformly distributed on the message set $\calW$, we obtain
\begin{align}
p(y|W\neq i)= P_{Y_1|X}(y|x). \label{eqn:p_P}
\end{align}
%\red{Does \eqref{eqn:p_P} depend on $x$? RHS seems to depend on $x$ but LHS doesn't: ANSWER: Since P(y|W=i) and P(y|W\neq i) depend on $x,x'$. But we see that the condition $D(P_{Y_1|X}(\cdot|x)\|P_{Y_1|X}(\cdot|x'))=0$ means that $x,x'$ relate to each other$.
Hence, it is easy to show that 
\begin{align}
p(y)&= P_{Y_1|X}(y|x),\\
p_i(y)&=p_i, %\quad \forall y \in \calY_1, \forall i \in \calW.
\end{align}
%for any  $x\in \calX\setminus\{x'\}$ such that  $D(P_{Y_1|X}(\cdot|x)\|P_{Y_1|X}(\cdot|x'))=0$,  
for all $i\in\calW$  and $y\in \calY_1$. Therefore, we have
\begin{align}
\bbE\left[\ln \calH(W|Y_1^n)-\ln \calH(W|Y_1^{n+1})\,\big|\, Y_1^n\right]=0.
\end{align}

Now, we treat the remaining case where the relative entropy is positive. For any $x\in \calX$ there always exists an $x' \in \calX \setminus \{x\}$ such that $D(P_{Y_1|X}(\cdot|x)\|P_{Y_1|X}(\cdot|x'))>0$. By choosing that $x'$ as a fixed symbol satisfying the preceding condition,~\eqref{eq91new} becomes a strict  inequality. Therefore, $\beta_{ix}$ must be zero or one. Consequently, for all fixed $i\in\calW$, all the values of $\beta_{ix}$ for  all $ x\in \calX$ except for one are zero.  % I CHANGED THIS PART SLIGHTLY COMPARED WITH PREVIOUS RELATED TO YOUR CONCERNS

Similarly, for any $x \in \calX \setminus \{x'\}$ such that $D(P_{Y_1|X}(\cdot|x)\|P_{Y_1|X}(\cdot|x'))>0$, we have
\begin{align}
\label{eq92key}
\frac{\partial^2 F_i}{\partial \alpha_{ix}^2}&=\sum_{y \in \calY_1} \left(P_{Y_1|X}(y|x)-P_{Y_1|X}(y|x')\right)^2 \frac{[p(y)-p_i p(y|W=i)]^2}{p(y) p^2(y|W=i)} \nn\\*
&\qquad\times \left[1-\left(\ln \frac{p(y)}{p_i p(y|W=i)}\right)^{-1}+\left(\ln \frac{p(y)}{p_i p(y|W=i)}\right)^{-2}\right] > 0.
\end{align}
Consequently, either $\alpha_{ix}=0$ or $\alpha_{ix}=1, x\in \calX$. 

From~\eqref{eq76new},~\eqref{eq77new}, and~\eqref{eq78new} together with above results, we obtain
\begin{align}
\label{eq93new}
\bbE\left[\calH(W|Y_1^n)-\calH(W|Y_1^{n+1})|Y_1^n\right]\leq \max\left\{0, \max_{x,x'}\max_{\eta} \left\{\sum_{y \in \calY_1} p(y) \ln \frac{\eta \ln \eta}{f(y)\ln f(y)}\right\}\right\},
\end{align}
where $\eta \in \{p_1,p_2,\ldots,p_M\}$, $(x, x') \in \calX^2$ and
\begin{align}
\label{eq94new}
p(y)&=\eta  P_{Y_1|X} (y|x)+(1-\eta) P_{Y_1|X} (y|x'), \\
\label{eq95new}
f(y)&=\eta \frac{P_{Y_1|X}(y|x)}{p(y)}.
\end{align}
 We see from~\eqref{eq94new} and~\eqref{eq95new} that
\begin{align}
\sum_{y \in \calY_1} p(y) \ln \frac{\eta\ln \eta}{f(y)\ln f(y)}=\sum_{y \in \calY_1} p(y) \ln \left[\frac{p^2(y)}{P_{Y_1|X}(y|x)P_{Y_1|X}(y|x')}\right]+\sum_{y \in \calY_1} p(y)\ln \left[\frac{P_{Y_1|X}(y|x')\ln \eta}{p(y)\ln f(y)}\right].
\end{align}
Note that
\begin{align}
\frac{P_{Y_1|X}(y|x')}{p(y)}=\frac{1-f(y)}{1-\eta}.
\end{align}
It follows that
\begin{align}
\ln \left[\frac{P_{Y_1|X}(y|x')\ln \eta}{p(y)\ln f(y)}\right]&=\ln\left[\frac{(1-f(y))\ln \eta}{(1-\eta)\ln f(y)}  \right]\\
&=\left[\ln(1-f(y))-\ln(-\ln f(y))\right]-\left[\ln(1-\eta)-\ln(-\ln \eta)\right].
\end{align}
%Note that given $Y_1^n$ then $\eta \in \{p(1|Y_1^n), p(2|Y_1^n),\ldots,p(M|Y_1^n)\}$ which is a fixed vector. 
From~\eqref{eq95new}, we have
\begin{align}
\sum_{y \in \calY_1} p(y) f(y)=\sum_{y \in \calY_1} \eta P_{Y_1|X}(y|x)=\eta. \label{eqn:equals_eta}
\end{align} 
Combining with the fact that  the function $x\mapsto\ln(1-x)-\ln(-\ln x)$ is concave on $(0,1)$~\cite[pp.~424]{Burnashev80}, we obtain
%and from~\eqref{eq95new}, we 
%\begin{align}
%\sum_{y \in \calY_1} p(y) f(y)&=\sum_{y \in \calY_1} \eta P_{Y_1|X}(y|x)=\eta,  
%\end{align} 
%we obtain 
the following almost surely  %\red{why do you use almost surely? what's random about \eqref{eqn:app_jens}?}
\begin{align}
\sum_{y \in \calY_1} p(y) \left[\ln(1-f(y))-\ln(-\ln f(y))\right] \leq \ln(1-\eta)-\ln(-\ln \eta). \label{eqn:app_jens}
\end{align} 
Note that $p(y)$ and $\eta$ are random because they    depend on $Y_1^n$ which is also  random (cf.~Eqns.~\eqref{eq70} and~\eqref{eqn:piy}). 
%\red{initially you said that \eqref{eqn:app_jens} holds almost surely. but it is non random so I removed ``almost surely'': It is random since p(y) and eta depends on Y_1^n as you see from (106)-(110). Yes, the notation is not good since it does not express the dependence on Y_1^n, but I follows from Burnashev}  
This means that
\begin{align}
\label{eq103new}
\sum_{y \in \calY_1} p(y)\ln \left[\frac{P_{Y_1|X}(y|x')\ln \eta}{p(y)\ln f(y)}\right] \leq 0.
\end{align}
In addition, observe that
\begin{align}
&p(y) \ln \left[\frac{p^2(y)}{P_{Y_1|X}(y|x)P_{Y_1|X}(y|x')}\right]\nn\\*
&=\left(\eta P_{Y_1|X}(y|x)+(1-\eta) P_{Y_1|X}(y|x')\right)\ln \left[\frac{\left(\eta P_{Y_1|X}(y|x)+(1-\eta)P_{Y_1|X}(y|x')\right)^2}{P_{Y_1|X}(y|x) P_{Y_1|X}(y|x')}  \right] \\ 
&=\left(\eta P_{Y_1|X}(y|x)+(1-\eta)P_{Y_1|X}(y|x')\right)\ln \left(\eta P_{Y_1|X}(y|x)+(1-\eta) P_{Y_1|X}(y|x')\right) \nn\\
&\quad + \left(\eta P_{Y_1|X}(y|x)+(1-\eta) P_{Y_1|X}(y|x')\right)\ln\left[\eta \frac{P_{Y_1|X}(y|x)}{P_{Y_1|X}(y|x')}+(1-\eta)\ln\frac{P_{Y_1|X}(y|x')}{P_{Y_1|X}(y|x)}\right]\\
&\leq \left(\eta P_{Y_1|X}(y|x)+(1-\eta) P_{Y_1|X}(y|x')\right)\ln\left[\eta \frac{P_{Y_1|X}(y|x)}{P_{Y_1|X}(y|x')}+(1-\eta)\ln\frac{P_{Y_1|X}(y|x')}{P_{Y_1|X}(y|x)}\right]\\
&= \left(\eta \frac{P_{Y_1|X}(y|x)}{P_{Y_1|X}(y|x')} P_{Y_1|X}(y|x')+(1-\eta)\frac{P_{Y_1|X}(y|x')}{P_{Y_1|X}(y|x)}P_{Y_1|X}(y|x)\right)\ln\left[\eta\frac{P_{Y_1|X}(y|x)}{P_{Y_1|X}(y|x')}+(1-\eta)\ln\frac{P_{Y_1|X}(y|x')}{P_{Y_1|X}(y|x)}\right]\\
&\le \left(\eta \frac{P_{Y_1|X}(y|x)}{P_{Y_1|X}(y|x')} +(1-\eta)\frac{P_{Y_1|X}(y|x')}{P_{Y_1|X}(y|x)} \right)\ln\left[\eta \frac{P_{Y_1|X}(y|x)}{P_{Y_1|X}(y|x')}+(1-\eta)\ln\frac{P_{Y_1|X}(y|x')}{P_{Y_1|X}(y|x)}\right] \label{eqn:removable}\\
&\le \max\left\{0,\eta P_{Y_1|X} (y|x) \ln \left[\frac{P_{Y_1|X}(y|x)}{P_{Y_1|X}(y|x')}\right]+(1-\eta) P_{Y_1|X} (y|x') \ln \left[\frac{P_{Y_1|X}(y|x')}{P_{Y_1|X}(y|x)}\right]\right\}.\label{eqn:removable2}
\end{align} Here,  note that the inequality in~\eqref{eqn:removable}  can be removed if $\ln\big[\eta \frac{P_{Y_1|X}(y|x)}{P_{Y_1|X}(y|x')}+(1-\eta)\ln\frac{P_{Y_1|X}(y|x')}{P_{Y_1|X}(y|x)}\big] \leq 0$. Inequality \eqref{eqn:removable2} follows from the convexity of the function $x\mapsto x\ln x$ for $x>0$.

%\red{IS THIS EXPLANATION READLLY NECESARY? WHAT IF THE OPPOSITE INEQ HOLDS? THEN THE INEQ. in~\eqref{eqn:removable}  ALSO HOLDS RIGHT?: For the opposite holds, we see that $P_{Y_1|X}(y|x')\leq 1, P_{Y_1|X}(y|x) \leq 1$, so we can remove from~\eqref{eqn:removable} } 

Hence, we obtain
\begin{align}
&\sum_{y \in \calY_1} p(y) \ln \left[\frac{p^2(y)}{P_{Y_1|X}(y|x)P_{Y_1|X}(y|x')}\right] \nn \\
&\leq \max\left\{0,\eta \sum_{y\in \calY_1} P_{Y_1|X} (y|x) \ln \left[\frac{P_{Y_1|X}(y|x)}{P_{Y_1|X}(y|x')}\right]+(1-\eta) \sum_{y \in \calY_1} P_{Y_1|X} (y|x') \ln \left[\frac{P_{Y_1|X}(y|x')}{P_{Y_1|X}(y|x)}\right]\right\}\\
\label{eq111new}
&\leq B_1\quad \mbox{a.s.}
\end{align}
%\red{what is random in \eqref{eq111new}? why do you use a.s.?}  
From~\eqref{eq93new},~\eqref{eq103new}, and~\eqref{eq111new} we have~\eqref{eq66new} for $j=1$. We obtain the inequalities for $j\in \{2,3,\ldots,K\}$ analogously.
\end{IEEEproof}
\begin{lemma}
\label{lem6new}
For any $n\geq 0$ and $y \in \calY_j$ the following inequalities hold almost surely  (cf.~Definition \ref{def:info_q})
\begin{align}
\ln\calH(W|Y_j^n)-\ln\calH(W|Y_j^{n+1})\,\big|\,Y_j^n, \{Y_{j,n+1} = y\}  \leq \ln T_j, \label{eqn:difference_H}
\end{align} for all $j=1,2,\ldots,K$. The conditioning on  the random variable $Y_j^n$ and the event $\{Y_{j,n+1} = y\}$  means that the inequalities~\eqref{eqn:difference_H} hold almost surely  $Y_j^n$  (i.e., for all realizations of $Y_j^n$) for a fixed realization of $Y_{j,n+1}=y$. 
\end{lemma}
\begin{IEEEproof} This proof is on  Burnashev's argument in~\cite{Burnashev1976} with some additional arguments in the corresponding optimization problem to account for the fact that the transmitted signal at time $n+1$, i.e. $X_{n+1}$, depends on $W$ and all $Y_1^n, \ldots,Y_K^n$. Note the  inequality~\cite[pp.~264]{Burnashev1976}
\begin{align}
\frac{\sum_{i=1}^K \alpha_i}{\sum_{l=1}^K \beta_l} \geq \min_i \frac{\alpha_i}{\beta_i}, \quad \alpha_i,\beta_i \geq 0. 
\end{align}
Using the same notation as in Lemma~\ref{lem5new}  and the fact that the function $x\mapsto -x\ln x$ is concave, we have for any $y\in \calY_1$ that
\begin{align}
\psi(y)&:=\frac{\calH(W|Y_1^{n+1})|Y_1^n, \{Y_{1,n+1}=y\}}{\calH(W|Y_1^n)|Y_1^n}\\
&=\frac{-\sum_{i=1}^M p_i(y)\ln p_i(y)}{-\sum_{i=1}^M p_i \ln p_i}\\
&\geq \min_{i}\left[\frac{p_i(y)\ln p_i(y)}{p_i \ln p_i}\right].
\end{align}
%\red{where is the notation $X(y)$ used subsequently?}
It follows that 
%\red{rearranged the next equation slightly. CHECK: YES, THE REARRANGEMENT MAKES IT BETTER TO UNDERSTAND}
\begin{align}
-\ln \psi(y)&= \ln\calH(W|Y_1^n)-\ln\calH(W|Y_1^{n+1})\Big|Y_1^n, \{Y_{1,n+1}=y \} \label{eq164:lem7a}\\
\label{eq164:lem7}
&\leq \ln \left\{\max_i \left[\frac{p_i \ln p_i}{p_i(y)\ln p_i(y)}\right]\right\}.
\end{align}
Similarly to the argument in the proof of Lemma~\ref{lem5new}, we first disregard all other  constraints and consider the optimization (maximization) problem   in the $\{ \ldots \}$ in~\eqref{eq164:lem7}
%\begin{align}
%\max_i \left[\frac{p_i \ln p_i}{p_i(y)\ln p_i(y)}\right],
%\end{align}
subject to the constraints
%of~\eqref{eq129new} \red{I don't understand which optimization problem you're considering. Write it down explicityly just like what you did for $F_i $ in the previous Lemma} only with the following constraints for all $ i\in \{1,2,\ldots,M\}$.
\begin{align}
\sum_{x\in \calX} \alpha_{ix} &=1, \\
\sum_{x\in \calX} \beta_{ix} &=1, \\
\alpha_{ix}&\geq 0,\\
\beta_{ix}&\geq 0.
\end{align}
Note that we have
\begin{align}
\label{eq129new}
p_i(y)=\frac{p_i \sum_{x\in \calX} \alpha_{ix} P_{Y_1|X}(y|x)}{p_i \sum_{x\in \calX} \alpha_{ix} P_{Y_1|X}(y|x)+(1-p_i)\sum_{x\in \calX} \beta_{ix} P_{Y_1|X}(y|x)}.
\end{align}Define
\begin{equation}
\chi_{x,x',\eta}:= \frac{\eta P_{Y_1|X}(y|x)}{ \eta P_{Y_1|X}(y|x)+(1-\eta) P_{Y_1|X}(y|x')} 
\end{equation}
and 
\begin{align}
 A_{x,x',\eta}:= \frac{\eta \ln \eta}{ \chi_{x,x',\eta} \ln \chi_{x,x',\eta} } .
\end{align}
Using the same arguments as~Lemma~\ref{lem5new}, we can show that
\begin{align}
\label{eq171:lem7}
 \frac{p_i \ln p_i}{p_i(y)\ln p_i(y)} \leq \max\left\{0, \max_{0\leq \eta \leq 1} \max_{x,x' \in \calX} A_{x,x',\eta}\right\}.
\end{align}
%where
%\begin{align}
% A_{x,x',\eta}:= \frac{\eta \ln \eta}{ \chi_{x,x',\eta} \ln \chi_{x,x',\eta} } 
%\end{align}
%and
%\begin{equation}
%\chi_{x,x',\eta}:= \frac{\eta P_{Y_1|X}(y|x)}{ \eta P_{Y_1|X}(y|x)+(1-\eta) P_{Y_1|X}(y|x')}.
%\end{equation}
Now, if $P_{Y_1|X}(y|x')\geq P_{Y_1|X}(y|x)$, we have
\begin{align}
 \max_{0\leq \eta\leq 1} A_{x,x',\eta} =\frac{P_{Y_1|X}(y|x')}{P_{Y_1|X}(y|x)}.
\end{align}
If $P_{Y_1|X}(y|x')< P_{Y_1|X}(y|x)$, then by using the fact that for any $0\leq x\leq 1$ and $ 1\leq a\leq 1/x$ we have
\begin{align}
\frac{x\ln x}{(ax)\ln (ax)} \leq \frac{1-x}{1-ax},
\end{align} we obtain
\begin{align}
\max_{0\leq \eta \leq 1} A_{x,x',\eta} &\leq \max_{0\leq \eta \leq 1} \frac{1-\eta}{1-\chi_{x,x',\eta}}  \\
&=\max_{0\leq \eta \leq 1} \frac{\eta P_{Y_1|X}(y|x)+(1-\eta) P_{Y_1|X}(y|x')}{P_{Y_1|X}(y|x')} \\
\label{eq178:lem7}
&=\frac{P_{Y_1|X}(y|x)}{P_{Y_1|X}(y|x')}.
\end{align}
%\red{The result of this Lemma is followed. CAN YOU BE MORE EXPLICIT HERE?} 
Consequently, the conclusion of the lemma in~\eqref{eq66new} follows by combining~\eqref{eq164:lem7},~\eqref{eq171:lem7}, and~\eqref{eq178:lem7}.
%\red{VARIOUS EQUATIOSN ABOVE}.... 
\end{IEEEproof}
\begin{lemma} 
\label{lem7newest}
The following inequalities for each  $1\leq j\leq K$ hold almost surely
\begin{align}
\bbE\left[\left(\ln \calH(W|Y_j^n)-\ln \calH(W|Y_j^{n+1})\right)_a|Y_j^n\right]\leq \varphi(a) 
\end{align}
where 
\begin{align}
\varphi(a)&:=\max_{1\leq j\leq K}\left(\ln T_j\right)_a. \label{eqn:def_varphi}
\end{align}
Under the condition $B_{\mathrm{max}}<\infty$, $\varphi(a)=0$ for $a$ sufficiently large.
\end{lemma}
\begin{IEEEproof}
From   Lemma~\ref{lem6new}  we know that for any $n\geq 0$ and $y \in \calY_1$ we have the following inequalities
\begin{align}
\ln\calH(W|Y_1^n)-\ln\calH(W|Y_1^{n+1})\,\Big|\, Y_1^n, \{Y_{1,n+1}=y_1\} &\leq \ln T_1. 
\end{align}
Since $\ln T_1$ is non-negative and  using the fact that if $x\leq y$ and $y\geq 0$ we have $(x)_a \leq (y)_a$ for any $a \in \bbR$, we obtain
\begin{align}
\left(\ln\calH(W|Y_1^n)-\ln\calH(W|Y_1^{n+1})\right)_a \,\Big|\, Y_1^n, \{Y_{1,n+1}=y_1\} &\leq \left(\ln T_1\right)_a.
\end{align} Therefore, we have for any $a\in \bbR$
\begin{align}
&\bbE\left[\left(\ln\calH(W|Y_1^n)-\ln\calH(W|Y_1^{n+1})\right)_a \Big|Y_1^n\right]\nn\\*
&=\sum_{y\in \calY_1} \bbP(Y_{1,n+1}=y_1|Y_1^n) \left(\ln\calH(W|Y_1^n)-\ln\calH(W|Y_1^{n+1})\right)_a \, \Big| \, Y_1^n, \{Y_{1,n+1}=y_1\} \label{eqn:condition} \\
&\leq \sum_{y\in \calY_1} \bbP(Y_{1,n+1}=y_1|Y_1^n) \left(\ln T_1\right)_a  \\
&=\left(\ln T_1\right)_a,
\end{align}
where the conditioning on $\{ Y_{1,n+1}=y_1\}$ in~\eqref{eqn:condition} means that $Y_{1,n+1}$ in the term $\ln\calH(W|Y_1^{n+1})$ takes on the value $y_1$. 
%\red{should \eqref{eqn:condition} have $\{Y_{1,n+1}=y_1\}$? looks strange since  we have $\bbP( Y_{1,n+1}=y_1|...)$: ANSWER: $\left(\ln\calH(W|Y_1^n)-\ln\calH(W|Y_1^{n+1})\right)_a \Big|Y_1^n=y_1^n =g(Y_{1,n+1})$ }
Similarly, for the other $j = 2,\ldots, K$,  we have 
\begin{align}
\bbE\left[\left(\ln\calH(W|Y_j^n)-\ln\calH(W|Y_j^{n+1})\right)_a \Big|Y_j^n\right] \leq \left(\ln T_j\right)_a .
\end{align}
Recall  the definition of $\varphi$ in  \eqref{eqn:def_varphi}. We note that since $B_{\mathrm{max}} <\infty$, we have $P_{Y_j|X}(y|x) >0$  for all $x\in \calX$ and $ y \in \calY_j$ for all $ j=1,2,\ldots,K$. It     follows that $T_j <\infty $ for all $ j=1,2,\ldots,K$ and so $\varphi(a)=0$ for $a$ sufficiently large. This concludes the proof of the lemma.
\end{IEEEproof}
The converse part of Theorem \ref{thm:main_res} can be  stated  succinctly as follows.
\begin{lemma}
\label{conversethm}
The reliability function for a DM-BC with common message and   VLFT satisfies  
\begin{align}
E(R) &\leq \min_{1\leq j\leq K} B_j\left(1-\frac{R}{C_j}\right), \quad \forall R< \upC. \label{eqn:converse_lemma}
\end{align}
\end{lemma}
\begin{IEEEproof}
The proof is similar to Burnashev's arguments in~\cite{Burnashev1976} and~\cite{Burnashev80}. There are some subtle differences, hence for completeness, we provide the entire proof. Here, a combination of~\cite{Burnashev1976} and~\cite{Burnashev80} makes the proof that the sequences  $\xi^{(j)}_n$ (as defined in \eqref{eqn:xi_def}  in the following) are submartingales     simpler. It is enough to show that \eqref{eqn:converse_lemma} holds for $\bbP(\tau<\infty)=1$ and $B_{\mathrm{max}}<\infty$. Now, as in  Burnashev's arguments~\cite{Burnashev80}, we   consider the $K$ random sequences
\begin{align}
\xi^{(j)}_n :=\begin{cases} C_j^{-1}\calH(W|Y_j^n)+n,& \; \mbox{if}\; \;\;\calH(W|Y_j^n) \geq A_j,\\ B_j^{-1} \ln\calH(W|Y_j^n)+b+n,&\;\mbox{if} \;\;\; \calH(W|Y_j^n) \leq A_j \end{cases}.  \label{eqn:xi_def}
\end{align} where $A_j$ is the largest positive root of the following equation in $x$:
\begin{align}
\label{eqkey2000}
\frac{x}{C_j}=\frac{\ln x}{B_j}+b.
\end{align} For $b$ sufficiently large, we will show that the $K$ sequences $\xi^{(j)}_n$ respectively form submartingles with respect to the filtrations $\{\sigma(Y_j^n)\}_{n=0}^{\infty}$ for $j=1,2,\ldots,K$. Note that when $b$ sufficiently large, \eqref{eqkey2000} can be shown to have two distinct positive roots $a_j,A_j$ and that $A_j/a_j$ can be make arbitrarily large by increasing $b$~\cite[pp.~256]{Burnashev1976}.

Indeed, first we suppose that $\calH(W|Y_1^n) \leq A_1$. Then, we obtain
\begin{align}
\bbE\left[\xi_n^{(1)}-\xi_{n+1}^{(1)}|Y_1^n\right] &=-1+\bbE\Big[B_1^{-1} \ln \calH(W|Y_1^n) + b-(B_1^{-1}\ln \calH(W|Y_1^{n+1})+b)\bone\{\calH(W|Y_1^{n+1}) \leq A_1\}\nn\\
&\qquad -C_1^{-1}\calH(W|Y_1^{n+1})\bone\{\calH(W|Y_1^{n+1}) > A_1\}\Big|Y_1^n\Big]\\
&\le -1+B_1^{-1}\bbE\left[\ln\calH(W|Y_1^n)-\ln \calH(W|Y_1^{n+1})\,\big|\, Y_1^n\right] \label{eqn:lessA1a}\\
&\le  -1 +B_1^{-1} \times B_1 =0.\label{eqn:lessA1b}
\end{align} Here, \eqref{eqn:lessA1a} follows from the fact that $x/C_1 \geq (\ln x)/B_1 + b$ for $x \geq A_1$ and \eqref{eqn:lessA1b} follows from   Lemma~\ref{lem5new}.

Now, suppose that $\calH(W|Y_1^n) > A_1$. Let $a_1$ be the smaller of the two positive roots of~\eqref{eqkey2000}. Then, for $b$ sufficiently large we obtain
\begin{align}
&\bbE\left[\xi_n^{(1)}-\xi_{n+1}^{(1)}|Y_1^n\right]\nn\\*
&=-1+C_1^{-1}\bbE\left[\calH(W|Y_1^n)-\calH(W|Y_1^{n+1})|Y_1^n\right]\nn\\*
&\quad +\bbE\left[(C_1^{-1}\calH(W|Y_1^{n+1})-B_1^{-1}\ln \calH(W|Y_1^{n+1})-b)\bone\{\calH(W|Y_1^{n+1} \leq A_1\}|Y_1^n \right]\label{eqn:geA1a}  \\
&\le -1+C_1^{-1}C_1 + \bbE\left[(C_1^{-1}\calH(W|Y_1^{n+1})-B_1^{-1}\ln \calH(W|Y_1^{n+1})-b)\bone\{\calH(W|Y_1^{n+1} \leq A_1\}|Y_1^n \right]\\
&=\bbE\left[(C_1^{-1}\calH(W|Y_1^{n+1})-B_1^{-1}\ln \calH(W|Y_1^{n+1})-b)\bone\{\calH(W|Y_1^{n+1}) \leq A_1\}|Y_1^n \right]\\
&=\bbE\left[(C_1^{-1}\calH(W|Y_1^{n+1})-B_1^{-1}\ln \calH(W|Y_1^{n+1})-b)\bone\{\calH(W|Y_1^{n+1}) \leq a_1\}|Y_1^n \right]\nn\\*
&\quad + \bbE\left[(C_1^{-1}\calH(W|Y_1^{n+1})-B_1^{-1}\ln \calH(W|Y_1^{n+1})-b)\bone \{a_1<\calH(W|Y_1^{n+1}) \leq A_1\}|Y_1^n \right]\\
&\le \bbE\left[(C_1^{-1}\calH(W|Y_1^{n+1})-B_1^{-1}\ln \calH(W|Y_1^{n+1})-b)\bone \{\calH(W|Y_1^{n+1}) \leq a_1\}|Y_1^n \right]  \label{eq192new2017} \\
&\le B_1^{-1}\bbE\left[(\ln \calH(W|Y_1^n)-\ln \calH(W|Y_1^{n+1}))\bone \{\calH(W|Y_1^{n+1}) \leq a_1\}|Y_1^n \right] \label{eq193lan} \\
&\le B_1^{-1}\bbE\left[(\ln \calH(W|Y_1^n)-\ln \calH(W|Y_1^{n+1}))\bone \Big\{\ln \calH(W|Y_1^n)-\ln \calH(W|Y_1^{n+1}) >\ln\left(\frac{A_1}{a_1}\right)\Big\}\Big|Y_1^n \right]  \label{eqn:geA1b}  \\
&=B_1^{-1}\bbE\left[(\ln \calH(W|Y_1^n)-\ln \calH(W|Y_1^{n+1}))_{\ln\left(\frac{A_1}{a_1}\right)}\Big|Y_1^n \right]\\
&\leq B_1^{-1}\varphi\left(\ln\left(\frac{A_1}{a_1}\right)\right) \label{eqn:geA1c}\\
&= 0.  \label{eqn:geA1d}
\end{align} Here,  \eqref{eqn:geA1a} follows from   Lemma~\ref{lem3new},~\eqref{eq192new2017} follows from the fact that $C_1^{-1}\calH(W|Y_1^{n+1})\leq B_1^{-1}\ln \calH(W|Y_1^{n+1})+b$ when $a_1<\calH(W|Y_1^{n+1}) \leq A_1$,~\eqref{eq193lan} follows from the fact that if $\calH(W|Y_1^{n+1})\leq a_1$ and $\calH(W|Y_1^n) > A_1$ we have $C_1^{-1}\calH(W|Y_1^{n+1})-b\leq C_1^{-1} a_1 -b =B_1^{-1} \ln a_1 \leq B_1^{-1} \ln A_1 \leq B_1^{-1}\ln \calH(W|Y_1^n)$, \eqref{eqn:geA1b} follows from the assumption that $\calH(W|Y_1^n) > A_1$, and \eqref{eqn:geA1c}, \eqref{eqn:geA1d}  follow from the Lemma~\ref{lem7newest} and the fact that $A_1/a_1$ can be make arbitrarily large by increasing $b$. The above arguments leading to~\eqref{eqn:geA1d}  and~\eqref{eqn:geA1d} together with~\eqref{eqn:lessA1b} confirm that  $\xi^{(1)}_n$ forms a submartingale with respect to the filtration  $\{\sigma(Y_1^n)\}_{n=0}^{\infty}$. A completely analogous argument  goes through for $j=2,3,\ldots,K$.  %That completes the proof for this part for $j=1$. Similar proof for $j=2,3,\ldots,K$.

Now, since we know that
\begin{align}
\label{eqn:xi_limit}
\xi_0^{(1)}=\bbE[\xi_0^{(1)}] \leq \bbE[\xi_{n\wedge\tau}^{(1)}]\leq \limsup_{n\to \infty} \bbE[\xi_{n\wedge \tau}^{(1)}],
\end{align}
it follows that for $N$ sufficiently large we have %\red{should $\xi_0$ be $\xi_0^{(1)}$? and can you explain why $C_{1}^{-1} \ln M =\xi_0$ after \eqref{eqn:ach_rate_e}}
\begin{align}
C_{1}^{-1} \ln M&=\xi_0^{(1)} \label{eqn:ach_rate_a1} \\
&\leq \limsup_{n\to \infty} \bbE[\xi_{n\wedge \tau}^{(1)}]\\
&  \leq C_1^{-1}\limsup_{n\to \infty} \bbE\left[\calH(W|Y_1^{\tau_1\wedge n})\bone\{\calH(W|Y_1^{\tau_1\wedge n})\geq A_1\}\right]\nn\\
&\quad + \limsup_{n\to \infty} \bbE\left[\tau_1 \wedge n\right]  + \limsup_{n\to \infty} B_1^{-1} \bbE\left[\ln\calH(W|Y_1^{\tau_1\wedge n})\bone\{\calH(W|Y_1^{\tau_1\wedge n})\leq A_1\} \right]+b\label{eqn:ach_rate_a0} \\
&\leq C_1^{-1}\limsup_{n\to \infty} \bbE\left[\calH(W|Y_1^{\tau_1\wedge n})\right]\nn\\
&\quad + \limsup_{n\to \infty} \bbE\left[\tau_1 \wedge n\right]  + \limsup_{n\to \infty} B_1^{-1} \bbE\left[\ln\calH(W|Y_1^{\tau_1\wedge n})\bone\{\calH(W|Y_1^{\tau_1\wedge n})\leq A_1\} \right]+b\\
&\le  C_1^{-1}\limsup_{n\to \infty} \bbE\left[\calH(W|Y_1^{\tau_1\wedge n})\right]  + \limsup_{n\to \infty} \bbE\left[\tau_1 \wedge n\right]  + \limsup_{n\to \infty} B_1^{-1} \ln \bbE\left[\calH(W|Y_1^{\tau_1\wedge n}) \right]+b \label{eqn:ach_rate_a}\\
&=C_1^{-1}\bbE\left[\calH(W|Y_1^{\tau_1})\right]+\bbE\left[\tau_1\right]  +  B_1^{-1} \bbE\left[\ln\calH(W|Y_1^{\tau_1})\right]\\
&\le  C_1^{-1}[1+\rvP_{\rme}(R_N,N)\ln M] + \bbE\left[\tau_1\right]+ B_1^{-1} \ln [h(\rvP_{\rme}(R_N,N))+\rvP_{\rme}(R_N,N) \ln M]+b \label{eqn:ach_rate_b}\\
&= C_1^{-1}[1+\rvP_{\rme}(R_N,N)\ln M] + \bbE\left[\tau_1\right]+ B_1^{-1} \ln [-\rvP_{\rme}(R_N,N)\ln \rvP_{\rme}(R_N,N) \nn\\*
&\quad -(1-\rvP_{\rme}(R_N,N))\ln(1-\rvP_{\rme}(R_N,N))+\rvP_{\rme}(R_N,N) \ln M]+b\\
&\le   C_1^{-1}[1+\rvP_{\rme}(R_N,N)\ln M] + \bbE\left[\tau_1\right]+ B_1^{-1} \ln [-\rvP_{\rme}(R_N,N)\ln \rvP_{\rme}(R_N,N)+\frac{1}{e}+\rvP_{\rme}(R_N,N) \ln M]+b\label{eqn:ach_rate_c}\\
&= C_1^{-1}[1+\rvP_{\rme}(R_N,N)\ln M] + \bbE\left[\tau_1\right]+ B_1^{-1} \ln [-\rvP_{\rme}(R_N,N)\ln \rvP_{\rme}(R_N,N)+\rvP_{\rme}(R_N,N) \ln M]+O(1)\label{eqn:ach_rate_d} \\
&= C_1^{-1}[1+\rvP_{\rme}(R_N,N)\ln M] + \bbE\left[\tau_1\right]+ B_1^{-1}\ln \rvP_{\rme}(R_N,N) + B_1^{-1} \ln (\ln M-\ln \rvP_{\rme}(R_N,N))+O(1).\label{eqn:ach_rate_e}
\end{align} Here,  \eqref{eqn:ach_rate_a1} follows from~\eqref{eqn:xi_def} and $\calH(W|Y_1^{0})=H(W)=\ln M$, \eqref{eqn:ach_rate_a0} follows from~\eqref{eqn:xi_def} and~\eqref{eqn:xi_limit}, \eqref{eqn:ach_rate_a} follows from the fact that for any random variable $G$,  $\bbE[(\ln G) \bone\{G\leq g\}]\leq \ln \bbE(G)$ for $g \geq 1$ (which is assured by taking $b$ sufficiently large so $A_1$ eventually becomes larger than $1$),  %\red{are you guaratneed that the threshold $A_1$ is greater than 1?: YES, as $b$ increases, $A_1$ will increase to infinity (Burnashev)}
   \eqref{eqn:ach_rate_b}  follows from   Lemma~\ref{lem1newest},   \eqref{eqn:ach_rate_c}   follows from the fact that $-x\ln x\leq 1/e$ for $0\leq x\leq 1$, and \eqref{eqn:ach_rate_d}  follows from the fact that $B_1 <\infty$.

Therefore, we obtain  %\red{do you mean that $\tau_k$ is $\tau_1$; YES, I CORRECTED IT}
\begin{align}
\ln M  &\leq 1+\rvP_{\rme}(R_N,N)\ln M + C_1 \bbE\left[\tau_1\right]+ C_1 B_1^{-1}\ln \rvP_{\rme}(R_N,N) + C_1 B_1^{-1} \ln (\ln M-\ln \rvP_{\rme}(R_N,N))+O(1)\\
&\leq 1+\rvP_{\rme}(R_N,N)\ln M + C_1 N + C_1 B_1^{-1}\ln \rvP_{\rme}(R_N,N) + C_1 B_1^{-1} \ln (\ln M-\ln \rvP_{\rme}(R_N,N))+O(1).
\end{align}
A similar bound holds for the other branches indexed by $j=2,\ldots,K$. 
%Similarly, we also have
%\begin{align}
%\ln M &\leq 1+\rvP_{\rme}(R_N,N)\ln M + C_j N+ C_j B_j^{-1}\ln \rvP_{\rme}(R_N,N) \nn\\
%&\quad + C_j B_j^{-1} \ln (\ln M-\ln \rvP_{\rme}(R_N,N))+O(1), \quad j=1,2,\ldots,K.
%\end{align}
It follows that for all $j=1,2,\ldots,K$, we have 
\begin{align}
E(R)&\leq \liminf_{N\to \infty}\,\,-\frac{\ln \rvP_{\rme}(R_N,N)}{N}\\ 
&\leq \limsup_{N\to \infty}\,\,-\frac{\ln \rvP_{\rme}(R_N,N)}{N}\\ 
&\leq \limsup_{N\to \infty}\,\,B_j\left(1-\frac{R_N}{C_j}\right),\\ 
&=  B_j\left(1-\frac{\liminf_{N\to \infty} R_N}{C_j}\right),\\
&\leq B_j\left(1-\frac{R}{C_j}\right)
\end{align}
for all $R<C_j$. 
Therefore, we finally obtain \eqref{eqn:converse_lemma} as desired.  
%\begin{align}
%E(R) &\leq \min_{1\leq j\leq K} B_j\left(1-\frac{R}{C_j}\right),
%\end{align} for all $R< \upC$.
\end{IEEEproof}
Let us now say a few words about the novelties in the converse proof   vis-\`a-vis Burnashev's works in~\cite{Burnashev1976} and~\cite{Burnashev80}. In the original work  on DMCs with variable-length feedback by Burnashev~\cite{Burnashev1976}, he proved   Lemma~\ref{lem5new} for the case $K=1$ under the  assumption that $\calH(W|Y_1^n)$ is bounded. Hence, the construction of submartingles in Lemma~\ref{conversethm} was more complicated. More specifically, Burnashev needed to make of use~\cite[Lemma~5]{Burnashev1976}, and the constructed submartingale is a combination of two other submartingales in~\cite[Eqn.~(4.20)]{Burnashev1976}. This is meant  to account for the  constraint concerning the boundedness of $\calH(W|Y_1^n)$. In a later work for the related problem of sequential hypothesis testing~\cite{Burnashev80}, Burnashev proved a lemma similar to Lemma~\ref{lem5new} under no constraints on $\calH(W|Y_1^n)$. However, as we pointed out in the remark in~\eqref{eq47remark}, this direct proof does not lead to the desired result for our setting in which $K\geq 2$. We need to adapt and combine the  two different proof techniques in~\cite{Burnashev1976}  and~\cite{Burnashev80} to prove  Lemma~\ref{lem5new}.
 
\appendices

\section{Proof of Lemma \ref{lem1b}}\label{app:exp_rvs}

\begin{IEEEproof}
We use the same proof technique  as in~\cite[Lemma 8]{TruongTan16a}. In Lemma \ref{lem1b}, $K$ may be greater than or equal to $3$, so a direct application of~\cite[Lemma 8]{TruongTan16a} is cumbersome. However, since we are not seeking tight bounds on the second-order term in the asymptotic expansion of $\bbE(\max\{X_{1L},X_{2L},X_{3L},...,X_{KL}\})$ as in~\cite[Lemma 8]{TruongTan16a}, it is enough to show that if the following conditions hold
\begin{align}
\label{eq632017}
\bbE  ( X_{jL} )&= L +O(\sqrt{L}),\quad j=1,2,\quad\mbox{and}\\
\label{eq642017}
\var(X_{jL}) &= O(L),\quad j=1,2,
\end{align}
then, we have 
\begin{align}
\bbE(\max\{X_{1L},X_{2L}\}) &= L+O(\sqrt{L}),\\
\label{oldtech}
\var(\max\{X_{1L},X_{2L}\})&=O(L).
\end{align}
 This is because if the desired statement in  \eqref{eqn:exp_max} holds for two sequences of random variables, it will hold for three if $\var(\max\{X_1,X_2\})=O(L)$ since  $\max\{X_1,X_2,X_3\}=\max\{\max\{X_1,X_2\},X_3\} $. This argument obviously holds verbatim if we have $K$ sequences of random variables.   Now, observe that
\begin{align}
\label{eqoldkey}
\max\{X_{1L},X_{2L}\}&=\frac{1}{2}\left[X_{1L}+X_{2L}+|X_{1L}-X_{2L}|\right] .
\end{align}
Moreover, we have
\begin{align}
\bbE\left(|X_{1L}-X_{2L}|^2\right)+\bbE\left(|X_{1L}+X_{2L}|^2\right)&=2[\bbE(X_{1L}^2)+\bbE(X_{2L}^2)]\\
&=2\left[\var(X_{1L})+(\bbE [X_{1L}])^2+ \var(X_{2L})+(\bbE [ X_{2L}])^2\right]\\
&=2\left[O(L)+(\bbE [X_{1L}])^2+ (\bbE [X_{2L}])^2\right].
\end{align}
In addition, we also have
\begin{align}
\bbE\left(|X_{1L}+X_{2L}|^2\right)&\geq \left(\bbE [X_{1L}+X_{2L}]\right)^2.
\end{align}
Hence, we obtain
\begin{align}
\left(\bbE|X_{1L}-X_{2L}|\right)^2&\leq \bbE\left(|X_{1L}-X_{2L}|^2\right)\\
&\leq 2\left[O(L)+(\bbE [X_{1L}])^2+ (\bbE [X_{2L}])^2\right]-\left(\bbE[X_{1L}+X_{2L}]\right)^2\\
&= O(L)+(\bbE [X_{1L}]-\bbE [X_{2L}])^2\\
&=O(L)+(L+O(\sqrt{L})-L-O(\sqrt{L}))^2\\
\label{eqoldkey1}
&= O(L).
\end{align}
It follows from~\eqref{eq632017},~\eqref{eqoldkey}, and~\eqref{eqoldkey1} that
\begin{align}
\bbE\left[\max\{X_{1L},X_{2L}\}\right]&=\frac{1}{2}\bbE\left[X_{1L}+X_{2L}\right]+O(\sqrt{L})\\
&=L+O(\sqrt{L}).
\end{align}
Now, we estimate the variance as follows:
\begin{align}
\var(\max\{X_{1L},X_{2L}\})&=\bbE\left(\max\{X_{1L},X_{2L}\}-\bbE\left[\max\{X_{1L},X_{2L}\}\right]\right)^2\\
&= \frac{1}{2}\bbE\left(X_{1L}+X_{2L}+|X_{1L}-X_{2L}|-\bbE[X_{1L}+X_{2L}+|X_{1L}-X_{2L}|]\right)\\
\label{eqh1}
&=\frac{1}{2}\bbE\left[\left(X_{1L}-\bbE[X_{1L}]+X_{2L}-\bbE[X_{2L}]+|X_{1L}-X_{2L}|-\bbE[|X_{1L}-X_{2L}|]\right)^2\right]\\
&\le \frac{3}{2}\bbE\left[(X_{1L}-\bbE[X_{1L}])^2+(X_{2L}-\bbE[X_{2L}])^2+(|X_{1L}-X_{2L}|-\bbE[|X_{1L}-X_{2L}|])^2\right] \label{eqn:var1}\\
&=\frac{3}{2}\left[\var(X_{1L})+\var(X_{2L})+\var(|X_{1L}-X_{2L}|)\right]\\
&\leq \frac{3}{2}\left[\var(X_{1L})+\var(X_{2L})+\bbE(|X_{1L}-X_{2L}|^2)\right]\\
&\le\frac{3}{2}\left[O(L)+O(L)+O(L)\right]\label{eqn:var2}\\
\label{eq872017}
&=O(L).
\end{align} Here, \eqref{eqn:var1} follows from the Cauchy-Schwarz inequality and \eqref{eqn:var2} follows from~\eqref{eq642017} and~\eqref{eqoldkey1}. Since $\var(\max\{X_{1L},X_{2L}\})\geq 0$,   we obtain from~\eqref{eq872017} that
\begin{align}
\var(\max\{X_{1L},X_{2L}\})=O(L).
\end{align}
\end{IEEEproof}

\section{Proof of Lemma~\ref{lem1newest}}
\label{append}
\begin{IEEEproof} % The following is the Burnashev's proof~\cite[Lemma 1]{Burnashev1976}. For completeness and compatible notations, we introduce the proof again.
We have
\begin{align}
\bbE\left[\calH(W|Y_1^{n \wedge \tau_1}\right]&=\sum_{i=1}^n \bbE\left[\calH(W|Y_1^i)|\tau_1=i\right]\bbP(\tau_1=i) + \bbE\left[\calH(W|Y_1^n)|\tau_1>n\right]\bbP(\tau_1>n).
\end{align}
Using the fact that $\calH(W|Y_1^n)$ is almost surely bounded by $\ln M$, we have for two natural numbers $m<n$ that
\begin{align}
\left| \calH(W|Y_1^{n \wedge \tau_1})-\calH(W|Y_1^{m \wedge \tau_1})\right|& \leq  \bbE\left[\calH(W|Y_1^n)|\tau_1>n\right]\bbP(\tau_1>n) \nn\\
&\quad + \sum_{i=m+1}^n \bbE\left[\calH(W|Y_1^i)|\tau_1=i\right]\bbP(\tau_1=i)+\bbE\left[\calH(W|Y_1^m)|\tau_1>m\right]\bbP(\tau_1>m)\\
&\leq M \left[\bbP(\tau_1>n) + \sum_{i=m+1}^n\bbP(\tau_1=i)+ \bbP(\tau_1>m)\right]\\
&=2\bbP(\tau_1>m)\ln M \to 0, \quad \mbox{as}\quad m\to \infty,
\end{align} which yields that $\lim_{n\to \infty} \bbE\left[\calH(W|Y_1^{n \wedge \tau})\right]$ exists since $\bbR$ is complete. 

Define the error event
\begin{align}
\calE: =\big\{\hat{W}_1 \neq W\big\}.
\end{align}
By Fano's inequality we have %for $n$ sufficiently large \red{Fano is non-asymptotic. why do we need for $n$ suff large: YES, I WAS WRONG?}
\begin{align}
H(W|\hat{W}_1,\tau_1=n) &\leq h[\bbP(\hat{W}_1\neq W|\tau_1=n)]+ \bbP(\hat{W}_1\neq W|\tau_1=n) \ln (M-1)\\
&= h[\bbP(\calE|\tau_1=n)]+ \bbP(\calE|\tau_1=n) \ln (M-1).
\end{align} 
Hence,
\begin{align}
H(W|\hat{W}_1,\tau_1=n) \leq h[\bbP(\calE|\tau_1=n)]+ \bbP(\calE|\tau_1=n) \ln (M-1).
\end{align}
It follows that
\begin{align}
\label{eq38}
\sum_{j=1}^M H(W|\hat{W}_1=j,\tau_1=n)\bbP(\hat{W}_1=j|\tau_1=n) \leq h[\bbP(\calE|\tau_1=n)]+ \bbP(\calE|\tau_1=n) \ln (M-1).
\end{align}
Now, for any random variable $Z$, define an $M$-tuple (vector)
\begin{align}
\bP_{W|Z=z}:= \left(P_{W|Z}(1|z), P_{W|Z}(2|z),\ldots, P_{W|Z}(M|z) \right). \label{eqn:ent_z}
\end{align}
In the following, we overload  the notation $\calH(\bP)$ to mean the entropy of the probability mass function defined by the vector $\bP$.  Observe that
\begin{align}
\label{eq40}
H(W|\hat{W}_1=j,\tau_1=n) =\calH( \bP_{W|\hat{W}_1=j,\tau_1=n} ), 
\end{align}
where $Z$ in \eqref{eqn:ent_z} is replaced by $(\hat{W},\tau_1)$ and $z$ by $(j,n)$. % \red{In~\eqref{eq40}, we have used overloaded the notation $\calH(\bP)$ to mean the entropy of the probability mass function defined by the vector $\bP$. CHECK} % CHECK: OK}
Now, we see that
\begin{align}
P_{W|\hat{W}_1,\tau_1}(w|j, n)&=\sum_{y_1^n} P_{W|Y_1^n \hat{W}_1,\tau_1}(w|y_1^n, j, n) P_{Y_1^n|\hat{W}_1,\tau_1}(y_1^n|j, n)\\
&=\sum_{y_1^n} P_{W|Y_1^n}(w|y_1^n) P_{Y_1^n|\hat{W}_1,\tau_1}(y_1^n|j, n) \label{eqn:mc3}\\
&=\bbE\left[P_{W|Y_1^n}(W=w|Y_1^n)\,\big|\,\hat{W}_1=j,\tau_1=n\right]. \label{eqn:exp_p}
\end{align} Here, \eqref{eqn:mc3} follows the Markov chain $W-Y_1^n - (\hat{W}_1,1\{\tau_1=n\})$.
%\red{changed \eqref{eqn:exp_p} slightly. CHECK: OK}    

In vector notation, \eqref{eqn:exp_p} means that
\begin{align}
\label{eq44}
\bP_{W|\hat{W}_1=j,\tau_1=n}=\bbE\left[ \bP_{W|Y_1^n} \,\big|\,\hat{W}_1=j,\tau_1=n\right],
\end{align}
where the expectation on the right is over the randomness of $Y_1^n$. 
Using~\eqref{eq40} and~\eqref{eq44} and Jensen's inequality   noting that $\bP\mapsto\calH(\bP)$ is concave, we have
\begin{align}
H(W|\hat{W}_1=j,\tau_1=n)&=\calH\left(\bbE\left[\bP_{W|Y_1^n}\,\big|\,\hat{W}_1=j, \tau_1=n\right]\right)\label{eq46a}\\
&\geq \bbE\left[\calH\left( \bP_{W|Y_1^n} \,\big|\, \hat{W}_1=j, \tau_1=n \right)\right].\label{eq46}
\end{align}
From~\eqref{eq38} and~\eqref{eq46} we obtain
\begin{align}
\sum_{j=1}^M \bbE\left[\calH\big(\bP_{W|Y_1^n}  \,\big|\,\hat{W}_1=j, \tau_1=n \big)\right] \bbP(\hat{W}_1=j|\tau_1=n) \leq h[\bbP(\calE|\tau_1=n)]+ \bbP(\calE|\tau_1=n) \ln (M-1).
\end{align}
Hence,
\begin{align}
\label{eq2702017}
\bbE\left[\calH\left(\bP_{W|Y_1^n}  \,\big|\,\tau_1=n \right)\right] \leq h[\bbP(\calE|\tau_1=n)]+ \bbP(\calE|\tau_1=n) \ln (M-1)
\end{align}
It follows that for $N$ sufficiently large
\begin{align}
\bbE[\calH(W|Y_1^{\tau_1})]&=\sum_{n=1}^{\infty} \bbE\left[\calH\left(\bP_{W|Y_1^n} \,\big|\,  \tau_1=n \right)\right] \bbP(\tau_1=n)\\
&\leq \sum_{n=1}^{\infty} \left[h[\bbP(\calE|\tau_1=n)]+ \bbP(\calE|\tau_1=n) \ln (M-1)\right] \bbP(\tau_1=n)\label{eq2412016b}\\
&\le h(\bbP(\calE))+\bbP(\calE) \ln (M-1),\label{eq2412016a}\\
&\le h(\rvP_{\rme}(R_N,N))+ \rvP_{\rme}(R_N,N) \ln (M-1).\label{eq2412016}
\end{align}
Here,  \eqref{eq2412016b} follows from~\eqref{eq2702017}, \eqref{eq2412016a} follows from the fact that the function $h(x)$ is concave and \eqref{eq2412016} follows from the increasing property of the entropy function $h(x)$ for $0\leq x\leq 1/2$,  $\calE \subset \cup_{j=1}^K \{\hat{W}_j \neq W\}$, and $\rvP_{\rme}(R_N,N)\to 0$ as $N \to \infty$ (so $\rvP_{\rme}(R_N,N)\le 1/2$ for $N$ sufficiently large). A completely analogous argument applies for $j=2,3,\ldots,K$.  % \red{why do we need that $\rvP_{\rme}(R_N,N)\to 0$?: Since we need $\rvP_{\rme}(R_N,N)\leq 1/2$, and when calculate reliability function, we need $\rvP_{\rme}(R_N,N)$ to zero }
%Similarly, we also have for all $j=2,3,\ldots,K$
%\begin{align}
%\bbE[\calH(W|Y_j^{\tau_j})] \leq h(\rvP_{\rme}(R_N,N))+ \rvP_{\rme}(R_N,N) \ln (M-1).
%\end{align}
\end{IEEEproof}
%\subsection*{Acknowledgements}
%  This work was supported by a Singapore Ministry of
%Education Tier 2 under grant  number R-263-000-B61-112.  
\bibliographystyle{unsrt}
\bibliography{isitbib}
\end{document}

%% file: preamble.tex
\usepackage[mathscr]{eucal}
\usepackage[cmex10]{amsmath}
\usepackage{epsfig,epsf,psfrag}
\usepackage{amssymb,amsmath,amsthm,amsfonts,latexsym}
\usepackage{amsmath,graphicx,bm,xcolor,url}
\usepackage[caption=false]{subfig} 
\usepackage{fixltx2e}%ordering of single and double column floats
\usepackage{array}%array and tabular environments
\usepackage{verbatim}
\usepackage{bm}
\usepackage{algorithmic, cite}
\usepackage{algorithm}
\usepackage{verbatim}
\usepackage{textcomp}
\usepackage{mathrsfs}
\usepackage{multirow}
\usepackage{epstopdf}
\catcode`~=11 \def\UrlSpecials{\do\~{\kern -.15em\lower .7ex\hbox{~}\kern .04em}} \catcode`~=13 

\allowdisplaybreaks[1]
 
\newcommand{\nn}{\nonumber}

\newcommand{\calE}{\mathcal{E}}
\newcommand{\calF}{\mathcal{F}}

\newcommand{\calH}{\mathcal{H}}
\newcommand{\calI}{\mathcal{I}}

\newcommand{\calT}{\mathcal{T}}

\newcommand{\calW}{\mathcal{W}}
\newcommand{\calX}{\mathcal{X}}
\newcommand{\calY}{\mathcal{Y}}
\newcommand{\calZ}{\mathcal{Z}}

\newcommand{\bP}{\mathbf{P}}

\newcommand{\rmA}{\mathrm{A}}

\newcommand{\rmc}{\mathrm{c}}

\newcommand{\rmd}{\mathrm{d}}

\newcommand{\rme}{\mathrm{e}}
\newcommand{\rmE}{\mathrm{E}}

\newcommand{\rmN}{\mathrm{N}}

\newcommand{\bbE}{\mathbb{E}}

\newcommand{\bbN}{\mathbb{N}}

\newcommand{\bbP}{\mathbb{P}}
\newcommand{\bbQ}{\mathbb{Q}}
\newcommand{\bbR}{\mathbb{R}}

\newcommand{\bbZ}{\mathbb{Z}}

\DeclareMathAlphabet{\mathbsf}{OT1}{cmss}{bx}{n}
\DeclareMathAlphabet{\mathssf}{OT1}{cmss}{m}{sl}% slanted sans serif

\newcommand{\rvP}{\mathsf{P}}

\DeclareSymbolFont{bsfletters}{OT1}{cmss}{bx}{n}  
\DeclareSymbolFont{ssfletters}{OT1}{cmss}{m}{n}
\DeclareMathSymbol{\bsfGamma}{0}{bsfletters}{'000}
\DeclareMathSymbol{\ssfGamma}{0}{ssfletters}{'000}
\DeclareMathSymbol{\bsfDelta}{0}{bsfletters}{'001}
\DeclareMathSymbol{\ssfDelta}{0}{ssfletters}{'001}
\DeclareMathSymbol{\bsfTheta}{0}{bsfletters}{'002}
\DeclareMathSymbol{\ssfTheta}{0}{ssfletters}{'002}
\DeclareMathSymbol{\bsfLambda}{0}{bsfletters}{'003}
\DeclareMathSymbol{\ssfLambda}{0}{ssfletters}{'003}
\DeclareMathSymbol{\bsfXi}{0}{bsfletters}{'004}
\DeclareMathSymbol{\ssfXi}{0}{ssfletters}{'004}
\DeclareMathSymbol{\bsfPi}{0}{bsfletters}{'005}
\DeclareMathSymbol{\ssfPi}{0}{ssfletters}{'005}
\DeclareMathSymbol{\bsfSigma}{0}{bsfletters}{'006}
\DeclareMathSymbol{\ssfSigma}{0}{ssfletters}{'006}
\DeclareMathSymbol{\bsfUpsilon}{0}{bsfletters}{'007}
\DeclareMathSymbol{\ssfUpsilon}{0}{ssfletters}{'007}
\DeclareMathSymbol{\bsfPhi}{0}{bsfletters}{'010}
\DeclareMathSymbol{\ssfPhi}{0}{ssfletters}{'010}
\DeclareMathSymbol{\bsfPsi}{0}{bsfletters}{'011}
\DeclareMathSymbol{\ssfPsi}{0}{ssfletters}{'011}
\DeclareMathSymbol{\bsfOmega}{0}{bsfletters}{'012}
\DeclareMathSymbol{\ssfOmega}{0}{ssfletters}{'012}

\newcommand{\eps}{\varepsilon}
\newcommand{\dotleq}{\stackrel{.}{\leq}}

\DeclareMathOperator*{\argmax}{arg\,max}

\DeclareMathOperator{\var}{\mathsf{Var}}

\newcommand{\bone}{\mathbf{1}}

\newtheorem{theorem}{Theorem} 
\newtheorem{lemma}{Lemma}

\newtheorem{corollary}{Corollary}
\newtheorem{definition}{Definition}

\newcommand{\qednew}{\nobreak \ifvmode \relax \else
      \ifdim\lastskip<1.5em \hskip-\lastskip
      \hskip1.5em plus0em minus0.5em \fi \nobreak
      \vrule height0.75em width0.5em depth0.25em\fi}